\newcommand{\f}{f}
\newcommand{\ie}{{\it i.e.}}
\newcommand{\diag}{{\rm diag}}
\newcommand{\be}{\begin{equation}}
\newcommand{\ee}{\end{equation}}
\newcommand{\bea}{\begin{eqnarray}}
\newcommand{\eea}{\end{eqnarray}}
\newcommand{\ba}{\begin{array}}
\newcommand{\ea}{\end{array}}
\newcommand{\al}{\alpha}
\newcommand{\ga}{\gamma}
\newcommand{\de}{\delta}
\newcommand{\ep}{\epsilon}
\newcommand{\si}{\sigma}
\newcommand{\Om}{\Omega}

\newcommand{\tr}{{\rm tr}}

\newcommand{\PP}{\mathrm{I}\kern -2.5pt \mathrm{P}}
\newcommand{\R}{\mathrm{I}\kern -2.5pt \mathrm{R}}
\newcommand{\Z}{\mathsf{Z}\kern -5pt \mathsf{Z}}
\newcommand{\C}{\mathsf{I}\kern -5pt \mathrm{C}}

\newcommand{\rar}{\rightarrow}
\newcommand{\non}{\nonumber}

\newcommand{\cN}{\mathcal{N}}
\newcommand{\cW}{\mathcal{W}}

\newcommand{\half}{{\textstyle {1\over 2}}}

\newcommand{\1}{1\kern -3pt \mathrm{l}}

\newcommand{\oddots}{ \begin{picture}(12,12) \put(2,2){\circle*{1.3}}
\put(7,6){\circle*{1.3}} \put(11,10){\circle*{1.3}}  \end{picture} }

\newcommand{\SU}{\mathrm{SU}}
\newcommand{\SO}{\mathrm{SO}}
\newcommand{\Sp}{\mathrm{Sp}}
\newcommand{\su}{\mathrm{su}}
\newcommand{\so}{\mathrm{so}}
\newcommand{\spl}{\mathrm{sp}}
\newcommand{\U}{\mathrm{U}}
\newcommand{\ul}{\mathrm{u}}

\newcommand{\orbgen}{\theta}
\newcommand{\Omp}{\Omega^\prime}
\newcommand{\e}{{\rm e}}
\newcommand{\one}{1}
\newcommand{\two}{2}
\newcommand{\three}{3}

\documentclass[12pt]{article}

\newdimen\tableauside\tableauside=1.0ex
\newdimen\tableaurule\tableaurule=0.4pt
\newdimen\tableaustep
\def\phantomhrule#1{\hbox{\vbox to0pt{\hrule height\tableaurule
width#1\vss}}}
\def\phantomvrule#1{\vbox{\hbox to0pt{\vrule width\tableaurule
height#1\hss}}}
\def\sqr{\vbox{%
  \phantomhrule\tableaustep

\hbox{\phantomvrule\tableaustep\kern\tableaustep\phantomvrule\tableaustep}%
  \hbox{\vbox{\phantomhrule\tableauside}\kern-\tableaurule}}}
\def\squares#1{\hbox{\count0=#1\noindent\loop\sqr
  \advance\count0 by-1 \ifnum\count0>0\repeat}}
\def\tableau#1{\vcenter{\offinterlineskip
  \tableaustep=\tableauside\advance\tableaustep by-\tableaurule
  \kern\normallineskip\hbox
    {\kern\normallineskip\vbox
      {\gettableau#1 0 }%
     \kern\normallineskip\kern\tableaurule}%
  \kern\normallineskip\kern\tableaurule}}
\def\gettableau#1 {\ifnum#1=0\let\next=\null\else
  \squares{#1}\let\next=\gettableau\fi\next}

\newcommand{\Yfund}{\tableau{1}}
\newcommand{\Ysymm}{\tableau{2}}
\newcommand{\Yasymm}{\tableau{1 1}}

\begin{document}

\begin{flushright} 
{\tt hep-th/0103047}\\ 
BRX-TH-484 \\ 
HUTP-00/A052 \\
BOW-PH-121  \\
\end{flushright}
\vspace{1mm} 
\begin{center}
{\bf\Large\sf 
Vacuum states of {\large $\cN=1^*$} mass deformations \\[-2 mm]
of {\large $\cN=4$} and {\large $\cN=2$} conformal gauge theories \\
and their brane interpretations 
}
\end{center}
\vskip 5mm 
\begin{center} 

Stephen G. Naculich\footnote{
Research supported in part by the NSF under grant no.~PHY94-07194 \\
\phantom{aaa}  through the ITP Scholars Program.}$^{,a}$, 
Howard J. Schnitzer
\footnote{Research supported in part by the DOE under grant
DE--FG02--92ER40706.}
${}^{\!\!\!,\!\!\!}$
\footnote{Permanent address.}$^{,b,c}$,
and Niclas Wyllard\footnote{
Research supported by the DOE under grant DE--FG02--92ER40706.\\
{\tt \phantom{aaa} naculich@bowdoin.edu; 
schnitzer,wyllard@brandeis.edu}\\}$^{,b}$ 
\end{center}
\vspace{.05in}

\begin{center}
$^{a}${\em Department of Physics\\
Bowdoin College, Brunswick, ME 04011}

\vspace{.2in}

$^{b,3}${\em Martin Fisher School of Physics\\
Brandeis University, Waltham, MA 02454}

\vspace{.2in}

$^{c}${\em Lyman Laboratory of Physics \\
Harvard University, Cambridge, MA 02138}
\end{center}
\vspace{.3in}

\vskip 2mm

\begin{abstract} 
We find the classical supersymmetric vacuum 
states of a class of $\cN = 1^*$ field theories 
obtained by mass deforming superconformal models with 
simple gauge groups and $\cN = 4$ or $\cN =2$  supersymmetry.
In particular, new classical vacuum states for mass-deformed 
$\cN = 4$ models with Sp($2N$) and SO($N$) gauge symmetry are found. 
We also derive the classical vacua 
for various mass-deformed $\cN=2$ models 
with Sp($2N$) and SU($N$) gauge groups 
and antisymmetric (and symmetric) hypermultiplets.
We suggest interpretations of the mass-deformed vacua 
in terms of three-branes expanded into five-brane configurations.
\end{abstract}

\newpage

\setcounter{equation}{0}
\section{Introduction}

Maldacena's conjecture relating string theories on 
anti-de Sitter (AdS) spaces 
to conformal field theories on their boundary \cite{Maldacena:1998}
has provided a powerful laboratory to study supersymmetric theories, 
particularly at strong 't Hooft coupling.  The best known example of 
the AdS/CFT
correspondence is the duality between 
four-dimensional $\cN = 4$ $\SU(N)$ Yang-Mills theory
and type IIB string theory on AdS$_5 \times S^5$.  
To make contact with more realistic and less constrained 
four-dimensional field theories, 
it is of interest to find examples of gauge/gravity duals with
reduced supersymmetry and broken superconformal invariance.
Orbifold theories that correspond to projections of the 
$\cN = 4$ SU($N$) theory by 
a discrete subgroup $\Gamma$ of the SU(4) R-symmetry group 
have been considered in refs.~\cite{Kachru:1998}.  
The resulting conformal field theories,
which are dual to IIB string theory on 
AdS$_5 \times S^5/\Gamma$,
have $\cN =$ 0, 1, or 2 supersymmetries, 
{and the} gauge group {is} generically of the form $\Pi_i \U(N_i)$.
Another class of models with an AdS/CFT correspondence are
orientifold models \cite{Kakushadze:1998a,
Fayyazuddin:1998,Witten:1998c,Aharony:1998,Gukov:1998b}. 
A detailed study of this correspondence
for IIB configurations consisting of D3 branes 
in the background of an orientifold plane, 
and in some cases, a $\Z_2$ orbifold and/or D7-branes, 
that give rise to four-dimensional $\cN =2$ (or $\cN =4$) 
gauge theories with at most two factors has been carried out
in ref.~\cite{Lozano:2000},
and the chiral primaries on the two sides of the correspondence matched.

Of particular interest for our present work are 
mass deformations of superconformal theories,
that break the supersymmetry to $\cN=1$.  
(These are referred to as $\cN=1^*$ theories.)
Such systems naively lead to naked singularities on the gravity side 
\cite{Girardello:1999}.  
However, a careful analysis by Polchinski and Strassler \cite{Polchinski:2000}
revealed that the naked singularity 
is actually replaced by an expanded brane source.  
Other works on $\cN=1$ gauge/gravity duals include \cite{Klebanov:1998}.

In this paper, we will focus on 
$\cN=1^*$ mass-deformations 
of four-dimensional superconformal gauge theories 
with simple gauge groups and 
$\cN = 4$ or $\cN=2$ supersymmetry.
The undeformed conformal theories arise 
as the low-energy effective theories on a stack of D3-branes 
in various orientifold and orbifold backgrounds
in type IIB string theory 
(see ref.~\cite{Lozano:2000} for a detailed description).
We will identify the classical vacuum states of these theories,
and propose corresponding D-brane interpretations.

In {section} 2,
we review the classical vacuum states of the 
mass-deformed $\cN = 4$ SU(N) Yang-Mills theory, \ie,
the $\cN = 1^*$ theory studied in ref.~\cite{Polchinski:2000}.
The massive classical vacuum states of the 
mass-deformed $\cN = 4$ Yang-Mills theories with 
Sp($2N$) and SO($N$) gauge groups were initially analyzed
and interpreted by Aharony and Rajaraman \cite{Aharony:2000}. 
In section 3,
we present new classical vacuum states 
(which preserve a subset of the gauge symmetry) for these theories,
and propose brane configurations 
corresponding to these new solutions.
Section 4 is devoted to the $\cN=1^*$ mass-deformation 
of the $\cN=2$ Sp($2N$) gauge theory 
with one hypermultiplet in the antisymmetric representation 
and four hypermultiplets in the fundamental representation,
which corresponds to D3-branes in the presence of an O7-plane
and D7-branes.
The vacuum states of this mass-deformed theory 
are given a brane interpretation.
In section 5 we study the mass-deformation of models
corresponding to D3-branes in a $\Z_2$-orbifold background
and an orientifold.  
The two resulting $\cN=2$ theories with simple gauge group are 
(i) SU($N$) with a hypermultiplet in the antisymmetric and 
symmetric representation,
and (ii) SU($N$) with two hypermultiplets in the antisymmetric
representation and four in the fundamental representation.  
The classical vacuum  states of these $\cN = 1^*$ theories are found,
and brane interpretations proposed.
{Our conventions and some technical details are collected in two appendices.} 

\setcounter{equation}{0}
\section{Review of the mass-deformed $\cN=4$ $\SU(N)$ model} \label{sectPre}

In this section, we review the 
mass-deformed $\cN=4$ SU($N$) gauge theory
known as the $\cN = 1^*$ theory \cite{Vafa:1994},
and the interpretation of its vacua in the brane picture \cite{Myers:1999}
and dual supergravity theory \cite{Polchinski:2000}.
The superpotential of the model, 
written in terms of $\cN=1$ superfields, is
\be \label{SUW} \cW =
\frac{2\sqrt{2}}{6g_{\mathrm{YM}}^2}\ep^{ijk} \tr
(\phi^i[\phi^j,\phi^k])\,,
\ee
where $\phi^i$  ($i = 1,2,3$) are  chiral superfields
transforming in the adjoint representation of SU($N$).
To this superpotential,
we add the mass deformation 
\be 
\label{SUDW}  
\cW_{\rm mass} =
\frac{\sqrt{2}}{g_{\mathrm{YM}}^2}\sum^3_{i=1} m_i \tr(\phi^i)^2\,,
\ee
which, when all the masses are non-zero,
breaks the supersymmetry down to $\cN=1$.
In this case, one can rescale the $\phi^i$'s to make the masses equal. 
In what follows, we concentrate on this case, 
denoting the common (real) mass by $m$.
The classical supersymmetric vacuum states 
are obtained by solving the F- and D-term equations, 
which are 
 \be
 \label{SUF} 
 {}[\phi^i,\phi^j] = - m \ep^{ijk}\phi^k \,,
 \ee
and
 \be 
 \label{SUD} 
 \sum_{i=1}^3 [\phi^i,(\phi^i)^{\dagger}]=0 \,,
 \ee
respectively.
Equation (\ref{SUF}) together with ({\ref{SUD}) 
imply that the $\phi^i$'s are anti-hermitian \cite{Kac:1999}.
The most general solution is 
\be
\label{suvacsoln}
\phi^i = m T^i\,,
\ee
where the $T^i$'s form an $N$-dimensional 
(in general reducible) representation 
of the $\su(2)$ Lie algebra (\ref{sutwo}). 
It is always possible to choose a block-diagonal basis,
 \be 
 \label{Ti}
 T^i = \left( \ba{ccc} T^i_{n_1}  & & \\ 
                        & \ddots &   \\
                        & & T^i_{n_l}
 \ea \right ),  \qquad \sum_{k=1}^{l} n_k = N \,,
 \ee
in which $T^i_{n_k}$ are the generators of the $n_k$-dimensional
irreducible representation of $\su(2)$ 
in standard form  (\ref{Tz}), (\ref{Txy}),
with  $T^\one=T^x$, $T^\two=T^y$, and $T^\three=T^z$.

The undeformed $\cN=4$ SU($N$) gauge theory
can be realized on a stack of $N$ D3-branes in a flat background. 
We take the D3-branes to span the 0123 directions. 
The positions of the D-branes in the transverse directions $z^i$
(where we define
$z^\three =  x^4 + i x^5$, $z^\one =  x^6 + i x^7$,  and
$z^\two =  x^8 + i x^9$)
are represented by the complex scalar fields $\phi^i$,
which are mutually commuting in the vacuum state of the undeformed theory.

In the mass-deformed theory, 
the D3-branes are polarized by the 
Myers effect \cite{Myers:1999} (see also \cite{Kabat:1998}). 
Since the generators $T^i_{n_k}$ satisfy the Casimir relation
\be 
\label{casimir}
(T^1_{n_k})^2 + (T^2_{n_k})^2 + (T^3_{n_k})^2 = c_2(n_k) \1_{n_k}\,,
\ee
where $c_2(n_k) = -\frac{1}{4}(n_k^2-1)$,
each block in the vacuum solution (\ref{suvacsoln}), (\ref{Ti})
can be interpreted as the equation for a non-commutative two-sphere
(in the 579 directions of the $\R^6$ transverse to the D3-branes).
In the dual $\mbox{\sl AdS}_5\times S^5$ supergravity theory,
it was shown \cite{Polchinski:2000} that the vacuum solution (\ref{Ti})
corresponds to the addition of a set of D5-branes to
the $\mathrm{AdS}_5\times S^5$ background, 
each with topology $\R^4\times S^2$, 
where the radius of the $k$th D5-brane is proportional to $n_k$, 
the dimension of the $k$th block in the field theory solution. 
(The supergravity approximation is valid provided all the $n_k$ are large.)

Support for this interpretation 
is obtained by comparing the gauge enhancement
on the two sides of the correspondence. 
On the field theory side, the unbroken gauge symmetry
for the solution (\ref{Ti}) {is generated by} all $\SU(N)$ matrices $U$
satisfying $U^{\dagger} \phi^i U = \phi^i$. 
When there are $m$ irreducible representations of the same dimension, 
one gets $\U(m)$ gauge enhancement; 
the total gauge enhancement is the product of all 
such factors divided by an overall $\U(1)$.
In particular,
when all the blocks are of different dimensions, 
the gauge symmetry is broken to an abelian subgroup.
This is in agreement with the supergravity side, 
in which each stack of $m$ coincident five-branes contributes   
an $\U(m)$ factor; 
the total gauge enhancement is the product of all 
such factors divided by an overall $\U(1)$.  
In ref.~\cite{Polchinski:2000}, 
the above heuristic arguments were firmly established 
by solving the supergravity field equations to first
order, in an expansion based on the smallness of the 
D5-brane charge relative to the D3-brane charge.

\setcounter{equation}{0}
\section{$\cN=4$ orientifold models:  $\Sp(2N)$ and $\SO(N)$ } \label{Ori}

There are two other classes of $\cN=4$ superconformal gauge theories
with simple gauge groups, namely $\Sp(2N)$ and $\SO(N)$.
These theories are realized on a stack of D3 branes in the
background of an O3 orientifold plane.
In this section we consider the $\cN=1^*$ mass deformation of these models,
first within the context of the supersymmetric field theory,
and then from the brane perspective.
These $\cN=1^*$ theories were studied
in the context of the AdS/CFT correspondence 
by Aharony and Rajaraman \cite{Aharony:2000}, 
who discussed the supergravity interpretation of some \cite{Kac:1999} 
(see also \cite{Hanany:1999}), 
but not all, of the gauge theory vacuum states.
We will exhibit a new class of gauge theory vacua,
and then discuss their brane interpretations.

\subsection{Field theory vacuum solutions} 

\subsubsection{$\cN=4$ $\Sp(2N)$} \label{N=4Sp}

The superpotential and mass deformation of the
$\cN=4$ Sp($2N$) gauge theory
are the same as in the SU($N$) theory  (\ref{SUW}), (\ref{SUDW}),
yielding identical F- and D-term equations
(\ref{SUF}), (\ref{SUD}). 
In addition, however, 
the $\phi^i$ must be $2N\times 2N$ matrices that satisfy 
 \be 
 \label{spdef} 
 J (\phi^i)^T J =  \phi^i \,,
 \ee
which defines the generators of the adjoint representation of $\Sp(2N)$.
The real matrix $J$ is the symplectic unit of $\Sp(2N)$, 
satisfying $J^T= -J$ and $J^2=-\1_{2N}$; the standard basis is
\be
J = \left( \ba{cc} 0 & \1_N \\
 		-\1_N  & 0  \ea \right ).
\ee

The general solution to the F- and D-term equations is, as before,
\be
\label{spvacsoln}
\phi^i = m T^i\,,
\ee
where the $T^i$'s generate a $2N$-dimensional 
(in general reducible) representation of $\su(2)$.
By a unitary transformation, $\phi^i \rar U \phi^i U^{\dagger}$, 
the $T^i$'s may be brought to block-diagonal form (\ref{Ti}). 
In this basis, the Sp($2N$) condition (\ref{spdef}) becomes 
\be
\label{spnewdef}
 g^* (\phi^i)^T g =  \phi^i
\quad\Rightarrow\quad
(\phi^i)^T g = - g \phi^i \,,
\ee
where $g= U^T J U$, and hence $g^T = - g$ and $g g^* =-\1_{2N}$. 
While the $\phi^i$ are block-diagonal matrices in this basis, 
the matrix $g$ need not be.

The classically {\it massive} vacua, 
\ie, those in which the gauge symmetry is completely broken,
were completely classified in ref.~\cite{Kac:1999},
and their dual supergravity interpretation found
in ref.~\cite{Aharony:2000}.
One classically massive vacuum was found for every partition
of $2N$ into {\it distinct,  even} integers.
In other words, the massive vacua correspond to solutions (\ref{spvacsoln}),
(\ref{Ti})
in which all the irreducible blocks have different, even dimensions.
One may understand this result  as follows:
first, 
breaking of the symmetry to an abelian subgroup
requires the irreducible blocks in $T^i$ to be distinct.
In this case, it may be shown (see {appendix \ref{appB}})
that the matrix $g$ must be block-diagonal.
With the irreducible representations written 
in the standard basis (\ref{Tz}), (\ref{Txy}), 
it may be further shown that 
(with the convention that $a_k = a_{-k}$ in (\ref{Txy})) 
within each block, $g$ is proportional to 
\be 
\label{K} 
K \; = \left( \ba{cccc}  &&& 1 \\
  		&&-1& \\
  		&1&& \\
  		\oddots &&& \\
\ea \right )\,,
\ee
with signs alternating down the reverse-diagonal,
and all other entries vanishing.
The matrix $K$ (and therefore $g$) is antisymmetric 
only when the dimension of the representation is even;
hence, all the irreducible blocks in (\ref{Ti}) 
must be even-dimensional.
These solutions break the gauge group completely,
as we will see below.

As noted in ref.~\cite{Aharony:2000}, the above solutions -- direct 
sums of distinct, even-dimensional representations of $\su(2)$ -- do 
not exhaust the classical vacua of the 
mass-deformed $\cN=4$ $\Sp(2N)$ theory;  
vacua in which the $\Sp(2N)$ gauge symmetry is only
partially broken are not included in this class.
We now present another class of classical vacua
of the Sp($2N$) gauge theory.

Consider a solution (\ref{spvacsoln}) 
in which two of the irreducible blocks
have the same (not necessarily even) dimension $n$,
\be
\label{spodd} 
T^i = \left( \ba{cc} T^i_n & 0 \\
 		0 & T^i_n  \ea \right ).
\ee
In this case, the part of the matrix $g$ acting on these two blocks 
need not be block-diagonal, but has the form
\be 
g = \left( \ba{cc}     g_{11} & g_{12} \\
 			-g_{12}^T & g_{22}
	\ea \right )\,,
\ee
where the $n{\times}n$-dimensional 
matrices $g_{ij}$ must each be proportional to $K$,
as shown {in appendix \ref{appB}}.
If $n$ is odd, then $g_{11}=g_{22}=0$, and by a phase transformation,
$g$ may put into the form
\be
\label{gform}
g = \left( \ba{cc} 0  & K \\
 	         -K^T & 0
		\ea \right ).
\ee
If $n$ is even, it is still possible to bring $g$ 
into the form (\ref{gform}) 
by a transformation that does not affect $T^i$.
Finally, $g= U^T J U$ is solved by $U = {\rm diag}(\1_n, K)$.
This allows the unitary transformation to be undone, 
and using the fact that $K  T_n^i K^T  =  -(T^i_n)^T$,
we obtain
\be
\phi^i = m \left( \ba{cc} T^i_n & 0 \\
 		          0 & (-T^i_n)^T \ea \right )\,,
\ee
which satisfies (\ref{spdef}), as may be directly verified.

Thus, in addition to the massive vacua discussed in ref.~\cite{Aharony:2000},
the mass-deformed \hbox{$\cN=4$} $\Sp(2N)$ gauge theory
has vacua corresponding to any pair of identical irreducible 
(not necessarily even-dimensional) su(2) representations.
If the representations are even-dimensional, 
then by a change of basis it is possible to diagonalize $g$,
\ie, put it in the form $\diag(K,K)$,
but for odd-dimensional representations, $g$ must act to 
exchange the pair.

There also exist vacua corresponding to solutions containing 
the direct sum of an arbitrary number $m$ 
of identical even-dimensional irreducible representations,
in which case $g$ can be put in the form $\diag(K,K, \cdots, K)$, 
and solutions comprising 
$2m$ identical odd-dimensional irreducible representations, 
in which case $g$ can be written  as the $m \times m$ unit matrix
tensored with (\ref{gform}).
The most general vacuum solution consists of a direct sum
of these two types of reducible representations,
of varying dimensions (see appendix \ref{appB} for more details).

\bigskip
\noindent{\it Unbroken symmetry group of the vacuum solutions}
\bigskip

Let us now discuss the gauge enhancements for each of the solutions above,
\ie, the subgroup of the original $\Sp(2N)$ that remains
unbroken by the scalar field vacuum expectation values
$\phi^i = m T^i$. 
We seek all matrices $U \in Sp(2N)$ 
(which obey $U^{\dagger}U=\1$ and $U^T J U = J$) 
such that $U^{\dagger} T^i U = T^i$. 
{Infinitesimally we obtain the conditions (writing $U=e^H$)} : 
$H^{\dagger} = -H$, 
$H^T J + J H=0$, 
and $[H,T^i]=0$.

First, if $T^i$ is an {\it irreducible} representation, 
then $[H,T^i]=0$ implies $H \propto \1$, by Schur's lemma. 
Combining this with $H^T J + J H=0$ shows that $H=0$;
thus, the gauge group is completely broken.

Next, let $T^i$ be reducible into {\it two} irreducible representations
\be 
\label{Titwo}
\left( \ba{cc} T^i_{n_1} & 0 \\
 	       0 & T^i_{n_2} 
\ea \right ).
\ee 
In the block-diagonal basis (\ref{Titwo}),
the condition  $J H^T J =H $ becomes 
\be
\label{spcondition}
g^* H^T g =H.
\ee
Write $H$ as $\pmatrix{ a & b \cr c & d}$.
If the irreducible representations are distinct
(and therefore, as shown above,  even-dimensional), 
then $[H,T^i]=0$ implies $b=c=0$ by Schur's lemma,
and $a$ and $d$ are proportional to the unit matrix.
The matrix $g$ is block-diagonal, with each block proportional to $K$.
Hence (\ref{spcondition}) implies that $a=d=0$, so again
there is no unbroken gauge symmetry.
Similarly, for the direct sum of an arbitrary number of {\it distinct} 
(even-dimensional) irreducible representations,
the gauge group is completely broken; 
all these solutions therefore correspond to massive vacua,
as asserted above.

If the two irreducible representations are identical 
(and have dimension $n$),
then Schur's lemma implies that $a$, $b$, $c$, and $d$ 
are all proportional to the unit matrix,
that is, $H = h \otimes \1_n$,
where $h$ is a $2 \times 2$ matrix.
The condition $H^\dagger = -H$ implies that $h$ is anti-hermitian.

If the two identical blocks are {\it odd-dimensional}, 
then $K^T = K$ and $K^2 = \1_n$,
so that $g  = i\sigma_y \otimes K$, 
using (\ref{K}) and (\ref{gform}).
Equation (\ref{spcondition}) then implies that
$ (i \sigma_y) h^T (i \sigma_y)  = h$,
that is, $h \in \spl(2)$.
Thus, two equal odd-dimensional blocks give rise to 
$\Sp(2)$ gauge enhancement. 
Generalizing to $2m$ identical odd-dimensional blocks,
one finds that $H = h \otimes \1_n$,
where $h$ is now a $2m\times 2m$-dimensional (anti-hermitian) matrix.
In this basis, $g$ can be written as $j \otimes K$
where $j$ is of the form $\1_m \otimes i\si_y$.  
Then (\ref{spcondition}) yields $j h^T j = h$,
that is, $h \in \spl(2m)$.
Hence, when $T^i$ consists of $2m$ identical blocks of odd dimension,
the unbroken gauge symmetry is $\Sp(2m)$.

If the two identical blocks are {\it even-dimensional},
then, by a change of basis that does not affect the form (\ref{spodd}), 
$g$ may be written $\1_2 \otimes K$,
where now $K^T= -K$ and $K^2 = -\1_n$.
Equation (\ref{spcondition}) then implies that $h^T = -h$,
\ie, $h\in\so(2)$.
Thus, with two identical even-dimensional blocks, 
the gauge enhancement is $\SO(2)$. 
The argument easily generalizes to $m$ identical even-dimensional blocks,
yielding SO($m$) gauge enhancement, 
as noted in ref.~\cite{Aharony:2000}.

In summary, for $2m$ odd-dimensional (resp.~$m$ even-dimensional) 
blocks of the same dimension,
the gauge enhancement is Sp($2m$) (resp.~SO($m$)). \footnote{In 
this paper, we are only discussing the classical vacua.
Quantum effects may cause a further splitting of the vacua
and consequent reduction of the gauge symmetry.}

\subsubsection{$\cN=4$ $\SO(N)$} \label{N=4SO}

The superpotential and mass deformation of the
$\cN=4$ SO($N$) gauge theory
are the same as in the SU($N$) theory  (\ref{SUW}), (\ref{SUDW}),
yielding identical F- and D-term equations (\ref{SUF}), (\ref{SUD}). 
In addition, however, the $N \times N$ matrices $\phi^i$ must satisfy 
 \be 
 \label{sodef} 
 (\phi^i)^T  =  -\phi^i \,,
 \ee
appropriate  to the generators of the adjoint representation of $\SO(N)$.

The general solution to the F- and D-term equations is, as before,
\be
\label{sovacsoln}
\phi^i = m T^i\,,
\ee
where the $T^i$'s generate a $N$-dimensional 
(in general reducible) representation of $\su(2)$.
By a unitary transformation, $\phi^i \rar U \phi^i U^{\dagger}$, 
the $T^i$'s may be brought to block-diagonal form (\ref{Ti}). 
In this basis, the SO($N$) condition (\ref{sodef}) becomes 
\be
\label{sonewdef}
 \f^* (\phi^i)^T \f =  - \phi^i
\quad\Rightarrow\quad
(\phi^i)^T \f = - \f \phi^i\,,
\ee
where $\f= U^T U$, and hence $\f^T = \f$ and $\f \f^* = \1_N$. 
While the $\phi^i$ are block-diagonal matrices in this basis, 
the matrix $\f$ need not be.

The classically {\it massive} vacua, 
\ie, those in which the gauge symmetry is completely broken,
were completely classified in ref.~\cite{Kac:1999},
and their dual supergravity interpretation found
in ref.~\cite{Aharony:2000}.
One classically massive vacuum was found for every partition
of $N$ into {\it distinct} odd integers.
In other words, the massive vacua correspond to solutions (\ref{sovacsoln}),
(\ref{Ti})
in which all the irreducible blocks have different, odd dimensions.
One may understand this result as follows:
first, breaking of the symmetry to an abelian subgroup
requires the irreducible blocks in $T^i$ to be distinct.
As in the previous section, it may be shown 
that the matrix $\f$ must be block-diagonal,
and within each block, $\f$ is proportional to $K$ (\ref{K}).
$K$ is symmetric only when the 
dimension of the representation is odd;
hence, all the irreducible blocks in (\ref{Ti}) 
must be odd-dimensional.
When the blocks are odd-dimensional, it
is possible to find a block-diagonal unitary matrix $U$ that rotates
the $T^i$'s in each block into real antisymmetric matrices
(\ie, $\so(N)$ matrices).
These solutions break the gauge group completely,
as we will see below.

As noted in \cite{Aharony:2000}, the above solutions 
-- direct sums of distinct, odd-dimensional representations of $\su(2)$ --
do not exhaust the classical vacua of the mass-deformed
$\cN=4$  $\SO(N)$ theory;  
vacua in which the gauge symmetry is only
partially broken are not included in this class.\footnote{
One example cited in \cite{Aharony:2000} 
is SO(6) $\to$ SU(2) $\times$ U(1).
}
We now present another class of classical vacua
of the SO($N$) gauge theory.

Consider a solution of the form   (\ref{spodd}) 
in which two of the irreducible blocks
have the same (not necessarily odd) dimension $n$.
The part of the matrix $\f$ acting on these two blocks 
need not be block-diagonal, but has the form
\be 
\f = \left( \ba{cc}     \f_{11}		& \f_{12} \\
 			\f_{12}^T	& \f_{22} \ea 
     \right )\,,
\ee
where the $n{\times}n$-dimensional 
matrices $\f_{ij}$ must each be proportional to $K$ (\ref{K}).
When $n$ is even, $f_{11}=f_{22}=0$, and by a phase transformation,
$f$ may be put into the form
\be
\label{fform}
\f = \left( \ba{cc} 0  & K \\
 	         K^T & 0
		\ea \right ).
\ee
When $n$ is odd, $f$ may still be brought into the form (\ref{fform}) 
by a unitary transformation that does not affect $T^i$.
Then, $\f= U^T U$ is solved by 
\be 
\label{USO} 
U = \frac{1}{\sqrt{2i}} 
      \left( \ba{cc} \1  & i K \\
 		     i K^T & \1
\ea \right )\,,
\ee
allowing the unitary transformation to be undone,  yielding
\be
\phi^i = {m\over 2} 
\left( \ba{cc} T^i_n-T_n^{iT}  &  -i   (T_n^i+T_n^{iT}) K \cr
            i K^T   (T_n^i+T_n^{iT}) &  T^i_n-T_n^{iT}   \ea \right ).
\ee
This is manifestly antisymmetric, in accord with (\ref{sodef}).
Specifically, we have
\be 
\label{soeven}
\phi^{1,3} = m \left( \ba{cc} 0  & -iT^{x,z}_n K \\
 		i K^T T^{x,z}_n  & 0
		\ea \right ) \; , \qquad
\phi^2 = m \left( \ba{cc} T^y_n  & 0 \\
 		0 & T^y_n
		\ea \right ).
\ee
Using (\ref{Tz}) and (\ref{Txy}), one sees that in this basis
the $\phi^i$ are all real and antisymmetric.

Thus, in addition to the massive vacua discussed in ref.~\cite{Aharony:2000},
the mass-deformed \hbox{$\cN=4$} $\SO(N)$ gauge theory
has vacua corresponding to any pair of identical irreducible 
(not necessarily odd-dimensional) su(2) representations.
If the representations are odd-dimensional, 
then by a change of basis it is possible to diagonalize $f$,
\ie, put it in the form $\diag(K,K)$,
but for even-dimensional representations, $f$ must act to 
exchange the pair.

There also exist vacua corresponding to solutions containing 
the direct sum of an arbitrary number $m$ 
of identical odd-dimensional irreducible representations,
in which case $f$ can be unitarily transformed
into the form $\diag(K,K, \cdots, K)$, 
and solutions comprising 
$2m$ identical even-dimensional irreducible representations, 
in which case $f$ can be written  as the $m {\times} m$ unit matrix
tensored with (\ref{fform}).
The most general vacuum solution consists of a direct sum
of these two types of reducible representations,
of varying dimensions.

\bigskip
\noindent{\it Unbroken symmetry group of the vacuum solutions}
\bigskip

Let us now discuss the gauge enhancements for the solutions above,
\ie, the subgroup of the original $\SO(N)$ that remains
unbroken by the scalar field vacuum expectation values
$\phi^i = m T^i$. 
The discussion is similar to that for Sp($2N$).
We seek all matrices $U \in SO(N)$ 
(which obey $U^T U=\1$ and $U^* = U$) 
that satisfy $U^T T^i U = T^i$. 
Writing $U=e^H$, these conditions become
$H^T = -H$, 
$H^* = H$, and $[H,T^i]=0$.

As before, if $T^i$ is irreducible, 
or reducible into {\it distinct} (odd-dimensional) representations, 
the symmetry group is completely broken.

If $T^i$ is reducible into two identical representations
of dimension $n$, 
then  $H = h\otimes \1_n$.
In the block-diagonal basis (\ref{spodd}),
the condition $H^T=-H$ becomes
\be
\label{socondition}
\f^* H^T \f = -H\,,
\ee
and $H^* = H$ transforms into a condition that,
together with eq.~(\ref{socondition}), 
implies that $h$ is anti-hermitian.

If the two identical blocks are {\it even-dimensional}, 
then $K^T = -K$ and $K^2 = -\1_n$,
so that $\f  = i\sigma_y \otimes K$,
by (\ref{K}) and (\ref{fform}).
Equation (\ref{socondition}) then implies that
$ (i \sigma_y) h^T (i \sigma_y)  = h$,
that is, $h \in \spl(2)$,
giving $\Sp(2)$ gauge enhancement. 
Generalizing to $2m$ identical even-dimensional blocks,
following the reasoning in section \ref{N=4Sp},
we find that the unbroken symmetry is $\Sp(2m)$.

If the two identical blocks are {\it odd-dimensional},
then, by a change of basis that does not affect the form (\ref{spodd}), 
$\f$ may be written $\1_2 \otimes K$,
where now $K^T=  K$ and $K^2 =  \1_n$.
Equation (\ref{socondition}) then implies that $h^T = -h$,
yielding SO(2) gauge enhancement.
Generalizing to $m$ identical odd-dimensional blocks,
we find that the unbroken symmetry is $\SO(m)$,
as noted in ref.~\cite{Aharony:2000}.

In summary, for $2m$ even-dimensional (resp.~$m$ odd-dimensional) 
blocks of the same dimension,
the gauge enhancement is Sp($2m$) 
(resp.~SO($m$)).${}^{5,}$\footnote{
The symmetry-breaking pattern mentioned in the previous footnote 
corresponds to taking two 2-dimensional and two 1-dimensional 
$\su(2)$ representations,
which yields the unbroken gauge group 
$\Sp(2) \times \SO(2) \cong \SU(2)\times \U(1)$.
}

\subsection{Brane interpretation of the vacua} \label{N=4dual}

The $\cN=4$ Sp($2N$) and SO($N$) gauge theories are 
realized as the low-energy effective theories on 
a stack of D3-branes probing a type IIB background with orientifold group
generated by $\Om^\prime = \Om(-1)^{F_L}R_{456789}$, 
where $\Om$ is the world-sheet parity operation, $(-1)^{F_L}$
reverses the sign of the left-moving Ramond sector,
and $R_{456789}$ changes the sign of the coordinates 
transverse to the D3-branes.
This corresponds to the presence of an orientifold 3-plane 
at the origin of the coordinates transverse to the D3-branes.

The bosonic fields of these models may be obtained as projections 
of the gauge fields $A^\mu$ and adjoint scalars $\phi^i$ 
of the $\cN=4$ $\SU(N)$ model as follows. 
Consider a stack of $N$ D3-branes in an 
orientifold background, where the orientifold group acts on the D3-branes 
via the $N \times N$ matrix $\gamma_{\Om^\prime}$ \cite{Gimon:1996a},
so that the fields are projected via
\be
A^\mu  = -\ga_{\Omp} (A^\mu)^T \ga_{\Omp}^{-1}, \qquad
\phi^i = -\ga_{\Omp} (\phi^i)^T \ga_{\Omp}^{-1}.
\ee
The two different gauge groups arise 
from different choices of how the NS-NS and R-R 3-forms 
are twisted under the orientifolding \cite{Witten:1998c}, 
and correspond to choices of discrete torsion.   
For Sp($N$) (where  $N$ must be even), $\gamma_{\Om^\prime} = J$,
leading to the restrictions (\ref{spdef}) on the adjoint scalars.
For SO($N$), $\gamma_{\Om^\prime} = \1$,
leading to the restrictions (\ref{sodef}) on the adjoint scalars.

First, consider the interpretation of the vacuum solutions 
of the mass-deformed theory
consisting of {\it distinct} irreducible representations. 
On the covering space of the orientifold,
each block of the solution corresponds 
to a two-sphere, as in the SU($N$) theory.
For these solutions, the projection matrix 
$g$ (or $f$) is block-diagonal,
so the orientifold acts within each block, 
geometrically projecting each $S^2$ to $\R\PP^2$.

Aharony and Rajaraman analyzed the dual supergravity theory
\cite{Aharony:2000}, 
with the near-horizon geometry $AdS_5\times \R\PP^5$,
in which these mass-deformed vacuum solutions
correspond to 5-branes of topology $\R^4\times\R\PP^2$, 
where the $\R\PP^2$ is embedded inside the $\R\PP^5$ factor,
and the $\R^4$ lies inside $AdS_5$.  
Supporting evidence for this interpretation was obtained
via an analysis of the D3-brane charges carried by the D5-branes,
and the fluxes of the gauge field and the NS-NS 2-form.

What is the brane interpretation of the 
other solutions (\ref{spodd}) found above? 
On the double cover, 
the pair of identical blocks corresponds to 
a pair of two-spheres with equal radius.

When the representations are odd-dimensional (even-dimensional)
in the $\Sp$ ($\SO$) gauge theory,
the action of the orientifold group 
necessarily exchanges the two blocks 
(see eqs.~(\ref{gform}) and (\ref{fform})),
so the pair of spheres are actually mirror images.
We conclude that our new solutions are to be interpreted as 
a single physical D5-brane with topology $\R^4\times S^2$,
\ie,  one D5-brane and its mirror.
In retrospect, the presence of such a solution is not surprising; 
the configuration it describes arises from a five-brane mirror pair 
away from the orientifold plane being brought to the orientifold plane. 
More generally, a solution containing $2m$ identical
odd-dimensional (even-dimensional) representations
corresponds to $m$ D5-branes,
whose associated $S^2$'s have equal radii,
and their mirrors.
As we saw above, the gauge enhancement for such a solution is $\Sp(2m)$.
This is analogous to the $\Sp(2m)$ gauge enhancement
of $2m$ coincident D3-branes, 
of which the first $m$ are mapped to the second $m$ 
under the orientifold,
that is, $\gamma_{\Om^\prime} = J$. 

When the representations are even-dimensional (odd-dimensional)
in the Sp (SO) gauge theory, 
the orientifold action 
in (\ref{gform}) and (\ref{fform}) still exchanges the two blocks,
so these solutions may also be thought of in terms of a brane-mirror 
pair of $S^2$'s. However, in this case $g$ and $f$ can also be 
transformed to the block-diagonal form $\diag(K,K)$. In this basis, 
each $S^2$ is mapped to itself by the orientifold symmetry, and we end 
up with a pair of $\R\PP^2$'s.
More generally, a solution containing $m$ identical
even-dimensional (odd-dimensional) representations
corresponds to $m$ $\R\PP^2$'s of equal radius.
As we saw above, the gauge enhancement for this solution
is $\SO(m)$.
This is analogous to the $\SO(m)$ gauge enhancement
of $m$ coincident {fractional} D3-branes, when each is mapped to itself 
under the orientifold symmetry ($\gamma_{\Om^\prime} = \1$). 
The two different descriptions discussed above 
are analogous to the usual transformation of two fractional branes 
into a single physical brane, {\it i.e.}, a brane-mirror pair.

\setcounter{equation}{0}
\section{$\cN=2$ orientifold model:
$\Sp(2N)$ + $\protect\Yasymm + 4~ \protect\Yfund$} \label{N=2Sp}

In this section, we turn to $\cN=2$ superconformal gauge theories
with a simple gauge group. 
There is only one such theory that arises from IIB string theory
in an orientifold (only) background, namely, 
$\cN=2$ Sp($2N$) gauge theory with one antisymmetric and 
four fundamental hypermultiplets \cite{Sen:1996}.
The field theory contains (in $\cN=1$ language) 
one vector multiplet, 
one hypermultiplet ${\phi_a}^b$ in the adjoint representation of $\Sp(2N)$, 
one hypermultiplet $A_{ab}$ in the antisymmetric representation, and 
one hypermultiplet $\tilde{A}^{ab}$ in the conjugate antisymmetric 
representation. 
In addition there are four hypermultiplets $Q_{a}^I$ 
in the fundamental representation,
and four hypermultiplets $\tilde{Q}^{Ia}$
in the conjugate fundamental representation.
Indices may be raised and lowered using the symplectic unit $J^{ab}$ 
and its inverse $J_{ab}$. 
Any representation is equivalent to its conjugate 
via the raising or lowering of indices using $J$;
for example, $\tilde{A}_{ab} = J_{ac} J_{bd} \tilde{A}^{cd} $
transforms in the antisymmetric representation. 
As a result of (\ref{spdef}),
the adjoint representation written with both indices down,
\ie, $\phi_{ab}= J_{bc} {\phi_{a}}^c$,
is symmetric.

The $\cN=2$ superpotential of this theory is 
\be 
\label{spantiW}
\cW = \frac{2\sqrt{2}}{g_{\mathrm{YM}}^2}
[ - 2\tilde{A}^{ab}{\phi_b}^c A_{ca} + \tilde{Q}^{aI}\phi_a{}^b Q^I_b ],
\qquad
\ba{rcl} 
a,b,c & = & 1 ~{\rm to} ~2N \,,
\\ 
I & = & 1 ~{\rm to}~ 4, 
\ea 
 \ee
to which we add the mass deformation
\be  
\label{spantiM}
\cW_{\rm mass} = \frac{\sqrt{2}}{g_{\mathrm{YM}}^2} 
\left\{ m [ {\phi_a}^b {\phi_b}^a  + A^{ab}A_{ab}
+ \tilde{A}^{ab}\tilde{A}_{ab}]  
+ M Q^I_a\tilde{Q}^{Ia}\right\}.
\ee
Finding the vacuum solutions will be facilitated by defining
\bea
\label{useful}
(\phi^\one)_a{}^b & = & J^{bc} A_{ac}, \hspace{1in}
   {s_a}^b =  Q^I_a\tilde{Q}^{Ib},                 \non\\
(\phi^\two)_a{}^b & = & J_{ac} \tilde{A}^{cb}, \hspace{1.05in} 
{t_a}^b =  \tilde{Q}^I_a {Q}^{Ib},                 \non\\
(\phi^\three)_a{}^b & = &\phi_a{}^b,
\eea
where, of course, $s$ and $t$ are not independent.
By virtue of the symmetry of $\phi_{ab}$,
the antisymmetry of $A_{ab}$ and $\tilde{A}^{ab}$,
and the definitions of ${s_a}^b$ and ${t_a}^b$,
these fields are constrained to satisfy
\bea
\label{spantidef}
 J (\phi^\one)^T J  & = &   -\phi^\one,    \hspace{1in}
 			J s^T J   =   t,  		\non\\
 J (\phi^\two)^T J  & = &  -\phi^\two,   \hspace{1in}
 			J t^T J   =   s, 		\non\\
 J (\phi^\three)^T J  & = &  \phi^\three.  
\eea

The superpotential (\ref{spantiW}) and mass deformation (\ref{spantiM}) 
may be rewritten in terms of the fields (\ref{useful}) as
\bea
\label{spantiWM}
\cW &=& 
\frac{2\sqrt{2}}{g_{\mathrm{YM}}^2}
\tr (\phi^1[\phi^2,\phi^3]    +  s \phi^3  ), \non\\
\cW_{\rm mass} &=&
\frac{\sqrt{2}}{g_{\mathrm{YM}}^2} \tr(m \phi^i \phi^i +  M s)\,,
\eea
giving rise to the F-term equations
\bea
\label{spantiF}
[\phi^\one, \phi^\two] &=& - m \phi^\three - \half (s+t) \,, \non  \\  
{}[\phi^\two, \phi^\three] & = & - m \phi^\one	\,,   \non \\  
{}[\phi^\three, \phi^\one] &=& - m \phi^\two\,,	\non\\  
{}[\phi^\three, s] &=& [\phi^\three,t] ~=~0 \,, \non\\  
\phi^\three s &=&  - \half M s \,, \non\\  
\phi^\three t &=& \half M t \,.
\eea
We now turn to the classical vacua of this theory
by solving the F-terms equations 
(\ref{spantiF}) subject to the restrictions (\ref{spantidef}).

\subsection{Field theory vacuum solutions}

\subsubsection{Case 1: $Q=\tilde{Q}=0$}
\label{Q=0}

Consider the branch of moduli space in which
the expectation values of the scalars in the fundamental hypermultiplets
$Q^{I}_a$ and $\tilde{Q}^{Ia}$ vanish (thus $s=t=0$).
The F-term equations (\ref{spantiF}) 
then reduce {to those encountered in previous sections}.
The general solution,
which also satisfies the D-term equations,
can be written as
\be
\label{spantivacsoln}
\phi^i = mT^i = m\left( \ba{ccc} T^i_{n_1}  & & \\ 
                        & \ddots &   \\
                        & & T^i_{n_l}
 		\ea \right ), 
\ee
where, as before, 
the $T^i_{n_k}$'s are generators of the irreducible representations of su(2).
In this block-diagonal basis, the constraints (\ref{spantidef}) become
\be
\label{spantinewdef}
 g^* (\phi^\one)^T g  =   -\phi^\one, \qquad
 g^* (\phi^\two)^T g  =   -\phi^\two, \qquad
 g^* (\phi^\three)^T g =  \phi^\three,
\ee
where $g= U^T J U$, which is not necessarily block-diagonal.
We must determine {\it which} of the solutions 
(\ref{spantivacsoln}) satisfy the constraints (\ref{spantinewdef}),
which differ from the constraints (\ref{spnewdef}) 
of the $\cN=4$ $\Sp(2N)$ theory.

We first show that this theory has no {\it classically massive} vacua,
\ie, vacua in which the gauge symmetry is completely broken.
Such vacua would correspond to solutions (\ref{spantivacsoln})
in which all the irreducible blocks had different dimensions.
In this case, it may be shown 
(see appendix \ref{appB}) using (\ref{spantinewdef}) 
that $g$ must be block-diagonal,
and within each block, proportional to 
\be 
\label{K'}
K^\prime = \left( \ba{cccc}  &&& 1 \\
  		&&1& \\
  		&1&& \\
  		\oddots &&& \\
\ea \right ).
\ee
This matrix is symmetric,   however,
and so cannot be written as $g = U^T J U$,
which obeys $g^T = -g$.
Thus, no solution 
consisting of a direct sum of {\it distinct} irreducible 
representations is possible.
In particular, a single irreducible representation is not possible.

Consider a solution (\ref{spodd}) 
in which two of the irreducible blocks have the same dimension $n$.
It may be shown, using (\ref{spantinewdef}), 
that the part of the matrix $g$ acting on these two blocks has the form
\be
\label{gnewform}
g = \left( \ba{cc} 0  & K^\prime \\
 	         -K^\prime & 0
		\ea \right ).
\ee
Then $g= U^T J U$ is solved by $U = {\rm diag}(1, K^\prime)$,
allowing the unitary transformation to be undone. 
Using the fact that 
$K' T^z K' = (-T^z)^T$ and
$K' T^{x,y} K' = (+T^{x,y})^T$,
we obtain
\be
\label{spantipair}
\phi^\one = m \left( \ba{cc} T^x_n & 0 \\
 		          0 & (T^x_n)^T \ea \right ),\qquad
\phi^\two = m \left( \ba{cc} T^y_n & 0 \\
 		          0 & (T^y_n)^T \ea \right ), \qquad
\phi^\three = m \left( \ba{cc} T^z_n & 0 \\
 		          0 & (-T^z_n)^T \ea \right ),
\ee
which satisfies the constraints (\ref{spantidef}), as may be directly verified.

Thus, while a single irreducible representation of su(2), or a direct sum
of distinct irreducible representations, do not correspond to 
classical vacua of the $\cN=2$ Sp($2N$) gauge theory 
(with an antisymmetric hypermultiplet and four fundamental hypermultiplets), 
we have shown that representations corresponding to 
any pair of identical irreducible representations do.
More generally, representations consisting of a direct sum of
pairs of identical irreducible representations also correspond
to vacua of this theory.

\bigskip
\noindent{\it Unbroken symmetry group of the vacuum solutions}
\bigskip

We now turn to the gauge enhancements for the solutions above,
\ie, the subgroup of the original $\Sp(2N)$ that remains
unbroken by the scalar field vacuum expectation values
$\phi^i = m T^i$. 
The analysis is similar to that of section \ref{N=4Sp}.

Consider the solution
consisting of two irreducible representations of dimension $n$. 
Then the elements $U=e^H$ of Sp($2N$) 
that obey $U^\dagger T^i U = T^i$ 
are of the form $H= h \otimes \1_n$, where $h^\dagger = -h$.
In the block-diagonal basis (\ref{spodd}),
the condition $J H^T J = H $ becomes 
\be
\label{spanticondition}
g^* H^T g =H\,,
\ee
where $g  = i\sigma_y \otimes K^\prime$ from (\ref{gnewform}).
Equations (\ref{K'}) and (\ref{spanticondition}) imply that
$ (i \sigma_y) h^T (i \sigma_y)  = h$,
that is, $h \in \spl(2)$,
giving $\Sp(2)$ gauge enhancement. 
Generalizing to $2m$ identical blocks,
and following the reasoning from section \ref{N=4Sp},
one finds that the unbroken symmetry is $\Sp(2m)$.
This result hold irrespective of whether the blocks
are even- or odd-dimensional.
Hence, as noted above, some part of the gauge symmetry
remains unbroken in all the classical vacua.\footnotemark[5] 

\subsubsection{Case 2: $A=\tilde{A}=0$}
\label{A=0}

Now consider the classical vacua for which the 
vevs of the antisymmetric fields vanish, $A = \tilde{A}=0$.
Setting  $\phi^\one=\phi^\two=0$ in (\ref{spantiF}),
we find that $\phi^\three$, $s$ and $t$ are mutually commuting, 
and may be simultaneously diagonalized.  
In this diagonal basis, we have
\be
\label{antinotzero}
\phi^\three = - {M \over 2}  \left( \ba{ccc} \1_r &    & \\
                                                 & - \1_r& \\
                                                 &       & 0\ea 
                               \right), \qquad
s = {Mm} \left( \ba{ccc} \1_r  &   &   \\
                                           & 0 &    \\
   					   &   & 0  \ea 
                         \right), \qquad
t = {Mm} \left( \ba{ccc} 0  &        &\\
                                         & -\1_r  &\\
          				 &        &0\ea 
                         \right),
\ee
where, by (\ref{spantidef}), we have
\be
 J = \left( \ba{ccc}
  0     & \1_r& 0 \\
  -\1_r & 0   & 0 \\
  0     & 0   & j \ea
	\right) .
\ee
The rank $r$ of the submatrices in (\ref{antinotzero})
must be less than or equal to four,
the number of fundamental hypermultiplets.
This is because the F-term equations
\bea
\phi_a{}^b \: Q^{~I}_b &=& - \half {M} \: Q^{~I}_a \non\\
\tilde{Q}^{Ia} \phi_a{}^b \: &=& - \half {M} \: \tilde{Q}^{Ib} ,
\eea
have at most four independent eigenvectors,
corresponding to $I=1$ to 4.
The maximal rank $r=4$ is obtained, for example, by setting
\bea
Q^{~I}_a  &\sim&  \left\{  \ba{c} \de^{~I}_a \\ 0 \ea \right. ,
\qquad
\tilde{Q}^{Ia}  \sim \left\{  \ba{c} \de^{Ia} \\ 0 \ea \right. 
\qquad {\rm for} \quad
\left\{ \ba{c}  a=1\; {\rm to} \; 4 \\
                a=5\; {\rm to} \; 8\ea \right.
\non\\
Q^{Ia} &=& J^{ab} Q_b^{~I}, \qquad
\tilde{Q}_a^{~I} = J_{ab} \tilde{Q}^{Ib}.
\eea

\subsubsection{Case 3: General case}
\label{generalcase}

We now consider the possibility 
that the scalar components of both antisymmetric
and fundamental hypermultiplet fields 
have non-vanishing vacuum expectation values.
Although we have not completely classified the solutions 
to the F-term equations in this case,
it seems likely that,
for generic values of $M$,
all the solutions decouple into direct sums of 
the solutions found in sections \ref{Q=0} and \ref{A=0}.
Since the blocks with $Q$ and $\tilde{Q} \neq 0$
were of rank eight at most,
they play a negligible role in the space of
classical vacuum solutions of the Sp($2N$) gauge theory 
in the large $N$ limit.
Since the large $N$ limit is a necessary condition 
for the validity of the dual supergravity theory,
we may neglect the fundamental hypermultiplets in this context, 
and focus on the vacuum solutions of sec.~\ref{Q=0}.

This neglect of the fundamental hypermultiplets
is consistent with the computation of the beta function 
for the Sp(2$N$) gauge coupling.  
The four fundamental hypermultiplets make a contribution 
to the beta function that is of order $1/N$ relative 
to that of the vector superfield 
and the antisymmetric hypermultiplet.
Thus, we may consistently neglect vacua
with non-zero vacuum expectation values of the scalar components
of $Q$ and $\tilde{Q}$ 
in the large $N$ limit.   

Similar arguments apply to the SU($N$) gauge theory
with two antisymmetric and four fundamental hypermultiplets
to be discussed in section \ref{SU2A}.

\subsection{Brane interpretation of vacua}
\label{N=2dual}

The $\cN=2$ Sp($2N$) gauge theory with one antisymmetric 
and four fundamental hypermultiplets
is realized as the low-energy effective theory on 
a stack of D3-branes probing a
type IIB background with orientifold group
generated by $\Om^\prime=\Om(-1)^{F_L}R_{45}$. 
This corresponds to the presence of an orientifold seven-plane
lying along the 01236789 directions. 
For consistency one needs to add a
stack of four D7-branes (and their mirrors)
parallel to the orientifold plane. 

The bosonic fields of this model 
(excluding those belonging to fundamental hypermultiplets) 
may be obtained as projections 
of the gauge fields $A^\mu$ and adjoint scalars $\phi^i$ 
of the $\cN=4$ $\SU(2N)$ model as follows. 
Consider a stack of  $2N$ D3-branes 
in an orientifold background, where the orientifold group acts 
via the $2N \times 2N$ matrix $\gamma_{\Omp}$.
The orientifold action projects the fields according to 
\be
\label{spantiprojection}
A^\mu  = - \ga_{\Omp}  (A^\mu)^T \ga_{\Omp}^{-1},
\qquad
\phi^{\one,\two} = + \ga_{\Omp} (\phi^{\one,\two})^T \ga_{\Omp}^{-1},
\qquad
\phi^\three = -\ga_{\Omp} (\phi^\three)^T \ga_{\Omp}^{-1}.
\ee
For this theory, $\gamma_{\Omp}=J$ \cite{Park:1998}, 
so the gauge group is projected to Sp($2N$),
and the projections (\ref{spantiprojection})
are identical to the field theory constraints (\ref{spantidef}).

This identification allows us to interpret the vacuum solutions 
in section \ref{Q=0} with D-brane configurations.
The lack of any vacuum solution corresponding to a 
single irreducible representation
(or to a direct sum of distinct irreducible representations)
corresponds in the brane interpretation to the fact
that the $\Z_2$ projection of a two-sphere  
(or a set of spheres of different radii)
by $\Om^\prime$
does not give a closed two-surface, on which
a D5-brane could be wrapped.

On the other hand, 
the solutions (\ref{spantipair}) consisting of a 
pair of identical (even- or odd-dimensional) irreducible su(2) representations
correspond to a pair of spheres of equal radius on
the covering space;
by virtue of (\ref{gnewform}),
these blocks are mapped to one another by the
orientifold symmetry,
so the pair of spheres are actually mirror images.
Thus, the vacuum solutions (\ref{spantipair}) 
correspond to a single physical D5-brane with 
topology $\R^4\times S^2$,
\ie,  one D5-brane and its mirror.
More generally, a solution containing $2m$ identical
even- or odd-dimensional representations
corresponds to $m$ D5-branes 
(whose associated spheres have equal radii) and their mirrors.
As we saw above, the gauge enhancement for such a solution is $\Sp(2m)$.

\setcounter{equation}{0}
\section{$\cN=2$ orbifold models with simple gauge groups} 
\label{Orb}

In this section, we consider mass deformations of
$\cN=2$ superconformal gauge theories
that arise from IIB string theory in {orientifolded} orbifold backgrounds.
Restricting our focus to models with a simple gauge group, 
there are only two such models \cite{Park:1998} (see also \cite{Erlich:1999}). 
These are SU($N$) with one antisymmetric and one symmetric hypermultiplet,
and SU($N$) with two antisymmetric and four fundamental hypermultiplets.
Both of these are orientifolds of $\Z_2$ orbifolds,
and both are examples of models ``without vector structure''
\cite{Witten:1998a,Park:1998}.

\subsection{Field theory vacuum states}

\subsubsection{$\cN=2$ orbifold model: 
$\SU(N) + \protect\Yasymm +\protect\Ysymm$}
\label{SUSA}

The $\cN=2$ SU($N$) gauge theory with hypermultiplets 
in the antisymmetric and symmetric representations
arises as the field theory on a stack of $2N$ D3-branes
in a type IIB background with orientifold group
$(1 + \orbgen)(1  + \al \Omp)$ \cite{Park:1998}.
Here $\orbgen= R_{6789}$ is the $\Z_2$  orbifold generator,
which acts as 
$z_\one \rar - z_\one$ and $z_\two \rar -z_\two$.
Also, $\Omp = \Om (-1)^{F_L}R_{45}$,
and $\al$ acts as $z_\one \rar i z_\one$ 
and $z_\two \rar -i z_\two$ (so $\al^2=\orbgen$). 

As in sections \ref{Ori} and \ref{N=2Sp}, 
the bosonic fields of the gauge theory may be obtained as projections 
of the gauge fields $A^\mu$ and adjoint scalars $\Phi^i$ 
of an $\cN=4$ $\SU(2N)$ model.
Consider a stack of $2N$ D3 branes in 
an orientifold background, where the elements $\orbgen$ and $\al \Om^\prime$ 
of orientifold group\footnote{The orientifold group is actually
generated by $\al \Om^\prime $ alone, since
$(\al \Om^\prime)^2 = \orbgen$ 
(which in terms of the $\ga$'s translates into 
$\ga_\theta = - \ga_{\al\Om'}(\ga_{\al\Om'}^T)^{-1}$), 
but it is useful to see the action of the orbifold generator separately.}
act on the D3-branes via the $2N \times 2N$ matrices \cite{Park:1998} 
\be 
\label{PUSA}
\ga_{\orbgen}  = i\left( \ba{cc}   \1_N & 0 \\
 				0  & -\1_N
	\ea \right ) \,, \qquad 
\ga_{\al\Omp}  = \e^{i\pi/4} \left( \ba{cc} 0  & \1_N \\
 				       i\1_N  & 0
\ea \right ) .
\ee 
The orbifold and orientifold projections 
impose the restrictions 
\be
\label{SUSAgaugegroup}
A^\mu  = \ga_{\orbgen}  A^\mu \ga_{\orbgen}^{-1},
\qquad
A^\mu  = - \ga_{\al\Omp}  (A^\mu)^T \ga_{\al\Omp}^{-1}\,,
\ee
on the SU($2N$) gauge field, projecting it to
\be 
\label{SUSAA} 
A^{\mu} = {\left( \ba{cc} a^{\mu} & 0 \\ 
			0 &    (-a^{\mu})^T \ea \right)} \,,\\ 
\ee 
where $a^\mu \in \U(N) \sim \SU(N) \times \U(1)$. 
(We ignore the $\U(1)$ factor,
which in any case is suppressed for large $N$.)
The scalar fields corresponding to D-brane {positions} 
are restricted to 
\bea
\label{SUSAproject}
\Phi^\one &=& - \ga_{\orbgen}    \Phi^\one    \ga_{\orbgen}^{-1}  ,\qquad
\Phi^\one  =  +i\ga_{\al\Omp} (\Phi^\one)^T \ga_{\al\Omp}^{-1}\,,\non\\
\Phi^\two &=& - \ga_{\orbgen}    \Phi^\two    \ga_{\orbgen}^{-1}  ,\qquad
\Phi^\two  =  -i\ga_{\al\Omp} (\Phi^\two)^T \ga_{\al\Omp}^{-1}\,,\non\\
\Phi^\three &=&  +\ga_{\orbgen}    \Phi^\three    \ga_{\orbgen}^{-1}  ,\qquad
\Phi^\three  =  - \ga_{\al\Omp} (\Phi^\three)^T \ga_{\al\Omp}^{-1}\,.
\eea
The projected $\Phi^i$'s become 
\be 
\label{Phis} 
\Phi^\one = \left( \ba{cc} 0 & S \\ -\tilde{A} & 0 \ea \right), \qquad 
\Phi^\two =\left( \ba{cc} 0 & A \\ \tilde{S} & 0 \ea \right),\qquad
\Phi^\three = \left( \ba{cc} \phi & 0 \\ 0 & -\phi^T \ea \right).
\ee 
Here $S$ ($A$) transforms in the symmetric $\Ysymm$
(antisymmetric $\Yasymm\,$) representation of $\SU(N)$;
$\tilde{S}$ and $\tilde{A}$ transform in the corresponding conjugate
representations. 

Using (\ref{Phis}), the superpotential 
\be
\label{SUSAW}
\frac{2\sqrt{2}}{g_{\mathrm{YM}}^2}
\tr (\Phi^1[\Phi^2,\Phi^3]),
\ee
inherited from the $\SU(2N)$ $\cN=4$ model, is projected to
\be 
\label{SUSAWW}
\cW =
\frac{4\sqrt{2}}{g_{\mathrm{YM}}^2}
\tr( \tilde{S}\phi S   +  \tilde{A}\phi A ) \; .
 \ee
The same projection 
when applied to the usual $\SU(2N)$ mass deformation (\ref{SUDW}),
however, 
gives vanishing masses for the $S$ and $A$ fields. 
Instead, we use the alternative mass deformation
\be
\label{SUSADW}  
\cW_{\rm mass} =
\frac{2 \sqrt{2}}{g_{\mathrm{YM}}^2}
\tr [  m_\phi (\Phi^\three)^2 +  m \Phi^\one \Phi^\two ],
\ee
which projects to
\be 
\label{SUSADWW}  
\cW_{\rm mass} = 
\frac{2\sqrt{2}}{g_{\mathrm{YM}}^2}
\tr( 2 m_{\phi} \phi^2 + m \tilde{S} S - m \tilde{A} A).
\ee
The projection of the deformation (\ref{SUSADW}) to 
the $\SU(N) + \Yasymm +\Ysymm$ theory
thus leads to the restriction $m_S=-m_A=m$, 
although in general the masses need not be related in this way.
The sum of the superpotentials (\ref{SUSAWW}) and (\ref{SUSADWW})
then gives rise to the F-term equations 
\bea
\label{SUSAF}
A \tilde{A} + S \tilde{S} & = & - 2 m_\phi \phi,  \non\\ 
\phi S + S \phi^T         & = & - m S,            \non\\
\phi A + A \phi^T         & = & + m A,            \non\\
\tilde{S} \phi  + \phi^T \tilde{S}  & = & - m \tilde{S}, \non\\
\tilde{A} \phi  + \phi^T \tilde{A}  & = & + m \tilde{A}.
\eea

\bigskip
\noindent{\it Vacuum solutions}
\bigskip

We now turn to finding solutions to the vacuum equations (\ref{SUSAF}).
First we solve the F-term equations of the fields $\Phi^i$ 
before imposing the projections (\ref{SUSAproject}).
If we redefine
\bea
\Phi^\one = -i(\phi^\one-i\phi^\two), \qquad
\Phi^\two = -i(\phi^\one+i\phi^\two), \qquad
\Phi^\three = -i\phi^\three, 
\eea
and set $m_\phi = m$,
then the superpotential (\ref{SUSAW}) and (\ref{SUSADW})
leads to the F-term equations
\be
{}[\phi^i,\phi^j] = - m \ep^{ijk}\phi^k \; .
\ee
The general solution is, as before,  $\phi^i = m T^i$.
Hence we may write
\be
\label{SUSAsoln}
\Phi^\one = -i m T^- = -i m (T^x - i T^y),\quad
\Phi^\two = -i m T^+ = -i m (T^x + i T^y),\quad
\Phi^\three = -i m T^z,
\ee
where the $T^i$ is any (in general reducible)
even-dimensional representation of the $\su(2)$ algebra.
Equation (\ref{SUSAsoln}) also satisfies the D-term equations
$\sum_{i=1}^3 [\Phi^i,(\Phi^i)^{\dagger}]=0$.

Next we determine {\it which} $\su(2)$ representations 
(\ref{SUSAsoln}) survive the projections (\ref{SUSAproject}),
\ie, can be cast into the form (\ref{Phis}).
Consider first the $2n$-dimensional irreducible 
representation $T_{2n}^i$.  
In the standard basis, (\ref{Tz}) and (\ref{Txy})
do not have the form (\ref{Phis}). 
However, if we relabel the rows and columns as follows:
$n+1 \to 1$,
$n+2 \to n+2$,
$n+3 \to 3$,
$n+4 \to n+4$,
up through $2n$, 
and
$n \to n+1$,
$n-1 \to 2$,
$n-2 \to n+3$,
$n-3 \to 4$,
down through $1$,
then the $\Phi^i$'s in (\ref{SUSAsoln}) 
are exactly of the form (\ref{Phis}) with
\bea
\label{SUSAirredsoln}
\phi &=& m  
\left( \ba{cccccc} 
    -\frac{1}{2} & & & & & \\
                 & \frac{3}{2} & & & & \\
  		 & & -\frac{5}{2} & & &  \\
		 & & & \frac{7}{2} & & \\
		 & & & &\ddots & \\
  		 & & & &       &(-)^n (n-\frac{1}{2}) \\
\ea \right )    \,,\non\\
S &=& 
2m \left( \ba{cccccc}
      a_0 &	 & 	& 	& 	&  \\
          &  0	 &a_{2} & 	& 	&  \\
          &a_{-2}& 0 	&   	& 	&   \\
  	  &	 &   	& 0 	& a_{4} &   \\
  	  & 	 & 	&a_{-4} & 0 	&  \\
  	  & 	 & 	& 	& 	& \ddots
\ea \right ) \,,\non\\
A &=& 
2m \left( \ba{cccccc}  0     & a_{-1}& 	& 	&	&\\
   		   a_{1} & 0     &    	&   	&	&\\
   		 	 &  	 & 0 	& a_{-3} 	&\\
   			 & 	 & a_{3}& 0  	&       &\\
   			 & 	 &      &    	& 0  &\\
   			 & 	 &      &    	&  &\ddots\\
\ea \right )\,,
\eea
where the $\su(2)$ commutation relations require
$a_k^2 = \frac{1}{4} (n^2-k^2)$.
The signs of the $a_k$ must be chosen to satisfy 
$a_{-k}= (-)^k a_{k}$
in order that $S$ and $A$ be symmetric and antisymmetric
respectively.  
This further implies that $\tilde{S}=S$ and $\tilde{A}=A$,
which is consistent with $\Phi^\two=(\Phi^\one)^\dagger$.
(Similarly, a representation consisting of a {\it direct sum} of 
such even-dimensional irreducible representations can
be written in the form (\ref{Phis}).)

Although a single irreducible odd-dimensional representation of
$\su(2)$ cannot be of the form (\ref{Phis}),
consider a solution (\ref{SUSAsoln}) in which
$T^i$ contains a pair of $d$-dimensional ($d = 2n+1$) irreducible 
representations in block-diagonal form
\be
\label{SUSAredform}
T^i=\left( \ba{ccc}  \raisebox{3pt}{$T^i_{2n+1}$}         &\vline& 
\raisebox{3pt}{0} 			\\
\cline{1-3} 
			\raisebox{-5pt}{0}  		&\vline & 
\raisebox{-5pt}{$(-T_{2n+1}^i)^T$ }
\ea \right ) \,,
\ee
where $T^i_{2n+1}$ are in the standard form (\ref{Tz}), (\ref{Txy}).
Specifically,
\bea
\label{SUSAredsoln}
\Phi^\three &=& m 
\left( \ba{ccccccccc} 
n	&  	 & 	&  	&\vline&	 &	 &	 &	 \\
 	& n-1	 & 	&  	&\vline&	 &	 &	 &	 \\
	&  	 &\ddots&  	&\vline&	 &	 &	 &	 \\
 	&  	 & 	& -n	&\vline&	 &	 &	 &	 \\
\cline{1-9}						
 	&  	 & 	&  	&\vline&-n	 &	 &	 &	 \\
 	&  	 & 	&  	&\vline&	 &-n+1	 &	 &	 \\
 	&  	 & 	&  	&\vline&	 &	 &\ddots &	 \\
 	&  	 & 	&  	&\vline&	 &	 &	 &n	 \\
\ea \right) \,,
\non\\
\Phi^\one  &=& 2m 
\left( \ba{ccccccccc} 
0	&  	 & 	&     &\vline	&	 &	 &	 &	 \\
a_{n-{1\over2}} &0& 	&     &\vline	&	 &	 &	 &	 \\
	&a_{n-{3\over2}}&0&   &\vline	&	 &	 &	 &	 \\
 	&  	 &\ddots&\ddots &\vline	& 	 &	 &	 &	 \\
\cline{1-9}
 	&  	 & 	&     &\vline   &\raisebox{-3pt}{0} & 
\raisebox{-3pt}{$-b_{n-{1\over2}}$} &	 &  \\
 	&  	 & 	&     &\vline	&  &0                &-b_{n-{3\over2}}&\\
 	&  	 & 	&     &\vline 	&  &		     & 0& \ddots \\
 	&  	 & 	&     &\vline	&  &	             &  & \ddots \\
\ea \right)   \,,
\non\\
\Phi^\two &=& (\Phi^\one)^\dagger\,,
\eea
where $a_k^2 = b_k^2 = \frac{1}{4} (n-k+\frac{1}{2})(n+k+\frac{1}{2})$
but the choice of signs is undetermined.
Clearly, (\ref{SUSAredsoln}) is not of the form (\ref{Phis})
but can be made so by relabeling the rows and columns as follows:
$1 \to 1$, 
$2 \to 2d-1$, 
$3 \to 3$, 
$4 \to 2d-3$,
up through $d \to d$, and 
$d+1 \to d+1$, 
$d+2 \to d-1$, 
$d+3 \to d+3$, 
$d+4 \to d-3$, 
up through $2d \to 2d$.
Then $\Phi^i$ will be exactly of the form (\ref{Phis}) with
\bea
\label{SUSAredmat}
\phi &=& m~ 
\left( \ba{cccccc} 
               n & & & & & \\
                 & n-1 & & & & \\
  		 & & n-2 & & &  \\
		 & & & \ddots & & \\
		 & & & & -n+1 & \\
  		 & & & &       &-n \\
\ea \right )   \,, \non\\
S &=& 2m  ~
\left( \ba{cccccc}
          &      & 		& 	& 	& 0        \\
          &  	 &              & 	& 0     & -b_{\frac{1}{2} - n}    \\
          &      & 	        & 0     & a_{n-\frac{3}{2}} &   \\
  	  &	 &     \oddots  &\oddots&       &          \\
  	  & 0	 & -b_{n-\frac{3}{2}} & & 	&          \\
  	0 & a_{\frac{1}{2}-n}	 & 	& 	& 	& 
\ea \right ) \,,\non\\
A &=& 2m ~
\left( \ba{cccccc}  
          &      & 		& 	& a_{n-\frac{1}{2}}	& 0   \\
          &  	 &              &-b_{\frac{3}{2}-n}& 0  &     \\
          &      & \oddots      &\oddots& 		&     \\
  	  &a_{n-\frac{3}{2}}& 0 &       &       	&     \\
-b_{n-\frac{1}{2}}& 0	 & 		&  	& 		&     \\
  	0 & 	 &      	& 	& 		& 
\ea \right )\,.
\eea
The choice of signs
$b_{n-\frac{1}{2}-k} = (-)^k a_{n-\frac{1}{2}-k}$
then makes $S$ symmetric and $A$ antisymmetric,
and also yields $\tilde{S}=S$ and $\tilde{A}=A$.

To conclude, we have exhibited two classes of classical vacua: 
those corresponding to any number of even-dimensional su(2) representations,
and those corresponding to pairs of identical odd-dimensional 
su(2) representations.

\vfil\newpage
\noindent{\it Unbroken symmetry group of the vacuum solutions}
\bigskip

Let us consider the gauge enhancement for each of the solutions
$\Phi^i$ described above.
As before, we seek traceless, antihermitian matrices $H$ 
that commute with $\Phi^i$,
and that satisfy the constraints (\ref{SUSAgaugegroup}) 
on the adjoint representation of the gauge group, namely
\be
\label{SUSAH}
H= \ga_{\orbgen}  H \ga_{\orbgen}^{-1},
\qquad
H= - \ga_{\al\Omp}  H^T \ga_{\al\Omp}^{-1}.
\ee
\ie, $H$ must take the form (\ref{SUSAA}).  

For a single even-dimensional representation,
Schur's lemma implies that $H \propto \1$,
and (\ref{SUSAH}) then implies that $H=0$;
the gauge symmetry is completely broken.
The same result holds for a direct sum of
{\em distinct} (even-dimensional) representations.

Consider $m$ identical $2n$-dimensional representations,
$T^i_{2n}\otimes\1_m$,
where $T^i_{2n}$ is related to $\Phi^i$ by (\ref{SUSAsoln}) and 
$\Phi^i$ is of the form (\ref{Phis}) with (\ref{SUSAirredsoln}).
Schur's lemma shows that $H = \1_{2n} \otimes h$,
where $h \in \ul(m)$.
Eq.~(\ref{SUSAH}) then implies that $h \in \so(m)$,
and the gauge enhancement is $\SO(m)$.

Next consider a pair of identical $(2n+1)$-dimensional representations,
$\1_2 \otimes T^i_{2n+1}$, 
where
$T^i_{2n+1}$  has the standard form (\ref{Tz}), (\ref{Txy}).
Schur's lemma implies $H= h\otimes \1_{2n+1}$, where $h \in \ul(2)$.
The unitary transformation 
\be
\label{unitary}
U = \left( \ba{cc} \1_{2n+1} & 0 \\
                  0      & K^T L
           \ea 
    \right)\,,
\ee
where $L = {\rm diag} (1,1,-1,-1,1,1,-1,-1,\ldots) $
puts the reducible representation into the form (\ref{SUSAredform})
(the presence of $L$ generates the relative signs
$b_{n-\frac{1}{2}-k} = (-)^k a_{n-\frac{1}{2}-k}$ 
between the two irreducible representations),
and puts $H$ into the form
\be
H = \left( \ba{cc}  h_{11} \1_{2n+1} & h_{12} K^T L \\
                    h_{21} L K & h_{22} \1_{2n+1}   \ea \right).
\ee
Finally, rearranging rows and columns as described just below 
eq.~(\ref{SUSAredsoln}) and then imposing eq.~(\ref{SUSAH}) yields 
$h_{12}=h_{21}= 0$ and
$h_{22}=-h_{11}$,
so that $h \in \ul(1)$.
Generalizing this procedure to $2m$ representations of dimension $(2n+1)$,
we find that $h \in \ul(m)$, so the unbroken gauge symmetry 
is $\U(m)$.

In summary, for $2m$ odd-dimensional 
(resp.~$m$ even-dimensional) 
blocks of the same dimension,
the gauge enhancement is U($m$) (resp.~SO($m$)).\footnotemark[5] 

\subsubsection{$\cN=2$ orbifold model: 
$\SU(N) + 2\, \protect\Yasymm + 4\, \protect\Yfund$}
\label{SU2A}

The $\cN=2$ SU($N$) gauge theory with two 
antisymmetric hypermultiplets 
and four fundamental hypermultiplets
arises as the field theory on a stack of D3-branes
in a type IIB background with orientifold group
$(1 +  \orbgen)(1  + \Omp)$ and
four D7-branes for consistency \cite{Park:1998}.
The analysis of this model is very similar to that in 
section \ref{SUSA}.
The bosonic fields of the gauge theory 
(excluding those corresponding to fundamental hypermultiplets) 
may be obtained as projections 
of the gauge fields $A^\mu$ and adjoint scalars $\Phi^i$ 
of the $\cN=4$ $\SU(2N)$ model.
Consider {a stack} of $2N$ D3 branes in {an orientifold background},
where the independent generators $\orbgen$ and $\Om^\prime$ 
of the orientifold group act on the D3-branes via the 
$2N {\times} 2N$  matrices 
\cite{Park:1998} 
\be 
\label{PU2A}
\ga_{\orbgen}  = i \left( \ba{cc}   \1_N & 0 \\
 					0  & -\1_N 
	\ea \right ) \,, \qquad 
\ga_{\Omp}  = \left( \ba{cc} 0  & \1_N \\
 			-\1_N  & 0
\ea \right ) .
\ee 
The orbifold and orientifold projections impose the restrictions 
\be
\label{SU2Agaugegroup}
A^\mu  = \ga_{\orbgen}  A^\mu \ga_{\orbgen}^{-1},
\qquad
A^\mu  = - \ga_{\Omp}  (A^\mu)^T \ga_{\Omp}^{-1}\,,
\ee
on the SU($2N$) gauge field, thus projecting it to 
\be 
\label{SU2AA} 
A^{\mu} = {\left( \ba{cc} a^{\mu} & 0 \\ 
			0 & (-a^\mu)^T \ea \right)}\,, \\ 
\ee 
where $a^\mu \in \U(N) \sim \SU(N) \times \U(1)$. 
(We ignore the $\U(1)$ factor,
which in any case is suppressed for large $N$.)
{The scalar} fields corresponding to D-brane {positions} are restricted to 
\bea
\Phi^\one &=& - \ga_{\orbgen}    \Phi^\one    \ga_{\orbgen}^{-1}  ,\qquad
\Phi^\one  =   +\ga_{\Omp} (\Phi^\one)^T \ga_{\Omp}^{-1} \,, \non\\
\Phi^\two &=&  -\ga_{\orbgen}    \Phi^\two    \ga_{\orbgen}^{-1}  ,\qquad
\Phi^\two  =  + \ga_{\Omp} (\Phi^\two)^T \ga_{\Omp}^{-1} \,, \non\\
\Phi^\three &=& + \ga_{\orbgen}    \Phi^\three    \ga_{\orbgen}^{-1}  ,\qquad
\Phi^\three  =   -\ga_{\Omp} (\Phi^\three)^T \ga_{\Omp}^{-1}\,.
\eea
The projected $\Phi^i$'s become
\be 
\label{SU2APhis}
\Phi^\one = \left( \ba{cc} 0 & A_2 \\
		-\tilde{A}_1 & 0 \ea \right)\,, \qquad 
\Phi^\two =\left( \ba{cc} 0 & A_1 \\ 
			\tilde{A}_2 & 0 \ea \right) \,,\qquad
\Phi^\three = \left( \ba{cc} \phi & 0 \\ 0 &
		-\phi^T \ea \right)\,.
\ee 
Here the $A_i$'s transform
in the antisymmetric $\Yasymm\,$ representation of $\SU(N)$ and
the $\tilde{A}_i$'s in the conjugate representation.

The projected fields obtained in this way 
do not include the hypermultiplet fields 
in the fundamental representation, 
whose scalar components could in principle 
have non-zero vacuum expectation values.
However, we expect that vacua with non-zero fundamental vevs 
will not be relevant in the large $N$ limit,
for reasons discussed in section \ref{generalcase}.
 
The superpotential for this model 
is exactly the same as in section \ref{SUSA},
namely, (\ref{SUSAW}), (\ref{SUSADW}),
with the addition of some terms for the fundamental hypermultiplets.  
The form of the mass deformation (\ref{SUSADW})
yields $m_2=-m_1= m$ for the masses of the two antisymmetric 
hypermultiplets, although this need not hold in general.
If we set the vacuum expectation values 
of the fundamental superfields to zero, 
the F-term equations for 
$\SU(N) + 2\,\Yasymm + 4\,\Yfund$ 
are given by (\ref{SUSAF}),
replacing $A$ with $A_1$ and $S$ with $A_2$.

\bigskip
\noindent{\it Vacuum solutions}
\bigskip

The analysis of the solutions to the F-term equations 
(with the fundamental superfield vevs vanishing)
is similar to that of section \ref{SUSA}.
Unlike the $\cN=2$ $\SU(N) + \Yasymm +\Ysymm$ theory,
a single even-dimensional irreducible representation
cannot be put into the form (\ref{SU2APhis}).
Essentially, this is because it is not possible to 
distribute an odd number of $a_n$'s in (\ref{Txy})
into two antisymmetric matrices. 
Similarly, no solutions consisting of {\em distinct} irreducible
representations  of su(2) is possible.
Thus, 
as in the $\Sp(N) + \Yasymm + 4 \Yfund$ theory,
there are no classically massive vacua.

Solutions (\ref{SUSAredform})-(\ref{SUSAredmat}) built
from pairs of odd-dimensional irreducible representations
{\it can} be put into the form (\ref{SU2APhis})
if we set $b_{n-\frac{1}{2}-k} = a_{n-\frac{1}{2}-k}$,
and replace $A$ with $A_1$ and $S$ with $A_2$.
This choice of signs also implies 
$\tilde{A_1}=A_1$ and $\tilde{A_2}=\,- A_2$.

Solutions can also be constructed from 
a pair of identical even-dimensional irreducible representations, as follows.
Consider the $2n$-dimensional representation
of the  $\SU(N) + \Yasymm + \Ysymm$ theory,
namely, eq.~(\ref{Phis}) with (\ref{SUSAirredsoln}),
but replacing $A$ with $S_1$ and $S$ with $S_2$.
Choosing $a_{-n}=a_{n}$ (which differs from section \ref{SUSA})
makes both $S_1$ and $S_2$ symmetric,
and implies $\tilde{S_1}=-S_1$ and $\tilde{S_2}=S_2$,
since $\Phi^\two=(\Phi^\one)^\dagger$.
Take the tensor product of this representation with $\1_2$,
and then perform the unitary transformation
$U=\mathrm{diag}(\1_{2n}, \1_n \otimes i\si_y)$
to obtain
\be
\label{SU2Apair}
\Phi^\one \!=\! \left( \ba{cc} 0 & S_2\otimes i\si_y \\ 
                \tilde{S_1}\otimes i\si_y & 0 \ea \right), 
\Phi^\two \!=\!\left( \ba{cc} 0 & S_1 \otimes i\si_y\\   
                 -\tilde{S_2}\otimes i\si_y & 0 \ea \right),
\Phi^\three \!=\! \left( \ba{cc} \phi\otimes \1_2 & 0 \\ 
                         0 & -\phi^T \otimes \1_2\ea \right).
\ee
These matrices {have} the correct form (\ref{SU2APhis})
for a vacuum solution for $\SU(N)$ with two antisymmetric hypermultiplets,
with $A_i = S_i \otimes i\si_y$.
This result generalizes to a solution
with any even number of identical even-dimensional representations.

To conclude, the vacua of 
the $\SU(N) + \Yasymm +\Ysymm$ theory
correspond to (a direct sum of)
pairs of identical irreducible representations of su(2);
the pairs can consist of either even- or odd-dimensional representations.

We observe that the relation between the solutions of the
$\SU(N) + 2\,\Yasymm +4\,\Yfund$ theory
and those of the 
$\SU(N) + \Yasymm +\Ysymm$ theory
{is analogous to} that between the
$\Sp(2N) + \Yasymm +4\Yfund$ theory
and the
$\Sp(2N) + \Ysymm$ (=adjoint) theory
(\ie, the pure $\cN=4$ $\Sp(2N)$ theory).

\bigskip
\noindent{\it Unbroken symmetry group of the vacuum solutions}
\bigskip

Let us consider the gauge enhancement for each the solutions
$\Phi^i$ described above.
Again, we seek traceless, antihermitian matrices $H$ 
that commute with $\Phi^i$,
and that satisfy the constraints (\ref{SU2Agaugegroup}) 
on the adjoint representation of the gauge group, namely
\be
\label{SU2AH}
H= \ga_{\orbgen}  H \ga_{\orbgen}^{-1},
\qquad
H= - \ga_{\Omp}  H^T \ga_{\Omp}^{-1}
\ee
\ie, $H$ must take the form (\ref{SU2AA}).  

Consider the vacuum solution consisting
of $2m$ identical $2n$-dimensional representations of su(2),
$T^i_{2n}\otimes\1_{2m}$,
where $T^i_{2n}$ is related to $\Phi^i$ by (\ref{SUSAsoln}) and 
$\Phi^i$ is of the form (\ref{Phis}) with (\ref{SUSAirredsoln}).
Schur's lemma implies that $H = \1_2 \otimes \1_{n} \otimes h$,
where $h \in \ul(2m)$.
The unitary transformation
$U=\mathrm{diag}(\1_{2mn},\1_n \otimes J)$,
where $J = i \si_y \otimes \1_m$,
puts $H$ into the form
\be
H= \left( \ba{cc}  \1_n \otimes h & 0			\\
			0         & \1_n \otimes  (-J h J)
          \ea \right).
\ee
Eq.~(\ref{SU2AH}) then implies that $JhJ = h^T$,
that is, $h  \in \spl(2m)$,
so that the unbroken symmetry group is $Sp(2m)$.

The analysis of the unbroken gauge symmetry of the solution 
consisting of $2m$ identical $(2n{+}1)$-dimensional representations
is completely analogous to subsection \ref{SUSA}, 
except that the unitary transformation (\ref{unitary})
does not include the matrix $L$ 
(since the relative sign between the irreducible representations 
is now $b_{n-\frac{1}{2}-k} = a_{n-\frac{1}{2}-k}$).
The conclusion is unaltered, 
and the unbroken gauge symmetry is $\U(m)$.

In summary, for $2m$ odd-dimensional 
(resp.~$2m$ even-dimensional) 
blocks of the same dimension,
the gauge enhancement is U($m$) (resp.~Sp($2m$)).\footnotemark[5]

\subsection{Brane interpretation of the vacua}

We now turn to the brane interpretations of the vacuum 
solutions found in subsections \ref{SUSA} and \ref{SU2A}. 
The coordinates on 
the cover space of the orientifolds  
are the $\Phi^i$'s 
(\ref{Phis}), (\ref{SU2APhis}) and are the most convenient 
coordinates to use when discussing the interpretation of 
the vacuum solutions. 

Before proceeding, however, let us also mention that 
a general method for dealing with branes at singularities 
was pioneered in ref.~\cite{Douglas:1996}.
This method is inherently field-theoretic 
and can be used to derive the nature of the singularity 
from the field theory alone.
Performing such an analysis for the model in section \ref{SUSA} 
(in the absence of the mass deformation) 
leads to the  invariant coordinates (in the 6789 directions) 
$x=-S\tilde{A}$, $y=A\tilde{S}$.
In the process one also introduces the auxiliary coordinate
$w=-A\tilde{A}$. 
The F-term equations imply that 
these coordinates satisfy the equation $x y = w^2$, 
which describes the $\Z_2$ orbifold singularity\footnote{To 
see this note that one may
parameterize the equation as $x=a^2$, $y=b^2$ and
$w=a b$. This parameterization has a redundancy since it is
unchanged under the 
$\mathsf{Z}\kern -4pt \mathsf{Z}_2$ 
transformation $(a,b)\rar -(a,b)$.}. 
The relation between $x,y$ and the $\Phi^i$'s is as follows. 
For the 6789 directions $\Phi^{1,2}$
transforms under the orbifold transformation as 
$\Phi^{1,2}\rar - \Phi^{1,2}$; 
thus, the invariant coordinates may be taken to be
$(\Phi^\one)^2$ and $(\Phi^\two)^2$. 
These are block-diagonal matrices
\be
\label{invariantmatrices}
(\Phi^\one)^2 = \left( \ba{cc} -S\tilde{A} & 0 \\ 
                         0  & -(-S\tilde{A})^T \ea \right), \quad 
(\Phi^\two)^2 =\left( \ba{cc} A\tilde{S} & 0 \\ 
                          0  & -(A\tilde{S})^T \ea \right)\,,
\ee 
in which the upper block corresponds to the physical branes and the
lower block corresponds to their orientifold mirrors. 
The invariant coordinates corresponding to the 6789 directions 
can thus be taken to be $x=-S\tilde{A}$ and $y=A\tilde{S}$,
the same expressions as above. 
In addition, for the 45 directions, 
$\phi$ may be interpreted as the invariant coordinate;
the lower block in $\Phi^\three$, $-\phi^T$, represents 
the coordinate of the mirror under the orientifold transformation 
and the orbifold generator does not act on the 45 directions. 
The virtue of the invariant coordinates is that the presence 
of the singularity is explicit. 

Let us now return to the interpretation of vacuum states in terms of 
D-brane configurations. 
The fields $\Phi^i$ that solve the F- and D-term equations 
can be transformed into block-diagonal form (\ref{Ti}), 
where each block satisfies the Casimir relation 
\be
\label{sphere}
\half (\Phi^\one \Phi^\two + \Phi^\two \Phi^\one ) +(\Phi^\three)^2 
= -m^2 \left[ (T^x)^2 + (T^y)^2 + (T^z)^2 \right]
= -m^2 c_2 (n) \1\,,
\ee
where $n$ is the dimension of the block. 
The hermitian matrices $X^i = - i T^i$ can be interpreted 
(for each block) as the coordinates for a 
non-commutative two-sphere of radius 
$ \sqrt{-c_2(n)}$ ($\sim  \half  n$ for large $n$) 
on the covering space.
 
The solution to the $\SU(N) + \Yasymm + \Ysymm$ gauge theory 
consisting of $m$ identical irreducible even-dimensional 
su(2) representations (\ref{SUSAirredsoln}) 
corresponds to $m$ copies of $S^2$ on the covering space.
The generators of the orientifold group act {\it within} each block
and consequently on the corresponding sphere, 
yielding $m$ copies of $S^2/\Z_4$. 
The brane picture is therefore analogous to
that for $m$ even-dimensional (odd-dimensional)
representations in the $\cN=4$ $\Sp(2N)$ ($\SO(N)$) gauge theories,
with the consequent $\SO(m)$ gauge enhancement.

The solution to the $\SU(N) + 2\,\Yasymm + 4\,\Yfund$ gauge theory 
consisting of $2m$ identical irreducible {\it even-dimensional} 
su(2) representations 
(direct sum of eq.~(\ref{SU2Apair}))
corresponds to $2m$ copies of $S^2$ on the covering space.
Consider this as $m$ pairs of spheres.
The orbifold generator $\theta$ 
(when transformed into the basis in which the solution is block-diagonal)
acts within each block,
but the generator $\Omega^\prime$ 
(similarly transformed)
exchanges the blocks within each pair.
Consequently, under the orientifold group projection,
we are left with $m$ copies of $S^2/\Z_2$.
The brane picture here is therefore analogous to
that for $2m$ odd-dimensional (even-dimensional)
representations in the $\cN=4$ $\Sp(2N)$ ($\SO(N)$) gauge theories,
with the consequent $\Sp(2m)$ gauge enhancement.

The solutions to the $\SU(N) + 2\,\Yasymm + 4\,\Yfund$ gauge theory
consisting of $2m$ identical irreducible {\it odd-dimensional}
su(2) representations (\ref{SUSAredform})-(\ref{SUSAredmat}) 
also corresponds to $2m$ copies of $S^2$ on the covering space.
Again, consider this as $m$ pairs of spheres.
In this case, however,
{\it all} of the (non-trivial) elements of the orientifold group
(when transformed into the basis in which the solution is block-diagonal)
act non-trivially on each pair 
(specifically, they act on the two-dimensional space as 
$\sigma_x$, $\sigma_y$ and $\sigma_z$)
as well as acting within each block.
The brane interpretation in this case is not completely transparent,
but observe that the orientifold group acts on the set of $2m$ D5-branes 
in the same way 
that the matrices $\gamma_\theta$, $\gamma_{\Omega^\prime}$ 
and $\gamma_{\theta} \gamma_{\Omega^\prime}$
defined in eq.~(\ref{PU2A}) act on $2N$ D3-branes,
leading to an analogous gauge enhancement:
in the latter case, $\U(N)$, and in the former, $\U(m)$.

A similar story holds for the solutions to the 
$\SU(N) + \Yasymm + \Ysymm$ gauge theory
consisting of $2m$ identical irreducible odd-dimensional
su(2) representations.

\section{Conclusions}

In this paper we have determined the classical vacuum states of 
$\cN=1^*$ mass deformations of 
$\cN=4$ and $\cN=2$ conformal models with simple gauge groups.
The undeformed theories arise as the effective field theories
on D3 branes in various orientifold and/or orbifold backgrounds.
The classical vacua were found from solutions
of the F- and D-term equations of the mass-deformed superpotentials, 
the gauge enhancements of these vacua were analyzed,
and interpretations in terms of D5-branes were suggested.  
These results are collected in Table 1,
which lists the building blocks of the vacuum solutions
of each of the gauge theories studied,
together with the resulting gauge enhancements 
and brane interpretation.
(The last column specifies the topology of the D5-branes, 
suppressing the $\R^4$ part.)

\begin{center} 
\begin{tabular}{|c|c|c|c|} 
\hline
Gauge theory & Vacuum solution & Enhancement & Interpretation\\ 
\hline
& & &  \\[-13pt]
\hline 
& & &  \\[-13pt]
$\Sp(2N)$ + adjoint  & $2m$ odd-dimensional irreps 
          & $\Sp(2m)$ & $m$ $S^2$'s \\[1pt]
\cline{2-4}
& & &  \\[-13pt]
{} & $m$ even-dimensional irreps & $\SO(m)$ & $m$ $\R\PP^2$'s \\
& & &  \\[-13pt]
\hline
& & &  \\[-13pt]
\hline
& & &  \\[-13pt]
$\SO(N)$ + adjoint  & $2m$ even-dimensional irreps 
           & $\Sp(2m)$ & $m$ $S^2$'s \\[1pt]
\cline{2-4}
& & &  \\[-13pt]
{} & $m$ odd-dimensional irreps & $\SO(m)$ & $m$ $\R\PP^2$'s \\
\hline
& & &  \\[-13pt]
\hline
& & &  \\[-13pt]
$\Sp(2N) + \Yasymm + 4 \Yfund$ & $2m$ irreps & $\Sp(2m)$ & $m$ $S^2$'s \\
& & &  \\[-13pt]
\hline
& & &  \\[-13pt]
\hline
& & &  \\[-13pt]
$\SU(N) + \Yasymm + \Ysymm$ & $2m$ odd-dimensional irreps 
& $\U(m)$ & see text \\[1pt]
\cline{2-4}
& & &  \\[-13pt]
{} & $m$ even-dimensional irreps & $\SO(m)$ &$m$ $S^2\!/\Z_4$'s \\[1pt]
\hline
& & &  \\[-13pt]
\hline
& & &  \\[-13pt]
$\SU(N) + 2 \,\Yasymm + 4\,\Yfund$ & $2m$ odd-dimensional irreps 
& $\U(m)$ & see text \\[1pt]
\cline{2-4}
& & &  \\[-13pt]
{} & $2m$ even-dimensional irreps & $\Sp(2m)$ & $m$ $S^2\!/\Z_2$'s \\[1pt]
\hline
\end{tabular}  
\end{center} 
\label{tablethree}
\centerline{\footnotesize{\bf Table 1}: 
Gauge enhancements and D-brane interpretations of $\cN=1^*$ vacua.}
\vspace{0.3cm}

The solutions consisting of $2m$ identical
odd-dimensional (resp. even-dimensional) 
representations of the $\cN=4$ Sp(2$N$) (resp. SO($N$)) 
gauge theories are new,
as well as the classification of vacua for 
various mass-deformed $\cN=2$ models with simple gauge groups.

The F-equations for all cases 
(with vanishing hypermultiplet fields in the fundamental representation) 
are equal to the F-equations of the 
$\cN=4$ theory with SU($N$) gauge group, 
supplemented by additional conditions specific to the
particular model being considered.
One therefore expects the first-order supergravity equations
to be given by projections of those of Polchinski and Strassler 
\cite{Polchinski:2000}.
In the large $N$ limit, 
we argue that hypermultiplets in the fundamental
representations essentially decouple and therefore may be neglected 
(see sections \ref{N=2Sp} and \ref{SU2A} for the relevant models).
Consequently, in the large $N$ limit, 
all the $\cN=1^*$ theories treated here allow a 
brane interpretation of the classical vacua in terms of D5 branes 
wrapped on closed two-surfaces.
Since the large
$N$ limit is a necessary condition for the supergravity dual, 
and hypermultiplets in the fundamental representation decouple in this
limit, 
the models studied here do not have a baryonic Higgs
phase in the dual supergravity description.

\section*{Acknowledgments}

We would like to thank Steve Gubser, Igor Klebanov, Joe
Polchinski, Matt Strassler, and Andy Waldron for conversations.

\newpage
\appendix
\section*{Appendices}
\setcounter{equation}{0}
\section{Some facts about $\su(2)$ representations} \label{su2app}

One choice for the commutation relations of 
the real Lie algebra $\su(2)$ is 
\be
\label{sutwo}
[T^i,T^j]= - \ep^{ijk}T^k.
\ee
The unitary representations of $\su(2)$ 
are given in terms of $n\times n$ anti-hermitian matrices, 
and there is one irreducible representation for each integer $n \ge 1$. 
A standard basis for the  generators of these representations is 
\be
\label{Tz} 
T_n^z = i\left( \ba{ccccc} j  & & & & \\
 			& j-1 & & &  \\
 			& & \ddots & & \\
 			& & & -j+1 & \\
 			& & & & -j
\ea \right ) \,,
\ee 
where $j$ is an integer or a half-integer, with $n=2j+1$. 
Both $T_n^x$ and $T_n^y$
have non-zero entries only directly above and directly below the
diagonal with $T_n^x$ symmetric (and hence purely imaginary) and
$T_n^y$ anti-symmetric (and hence real): 
\be 
\label{Txy} 
T_n^x = i\left( \ba{cccc} 0 & a_{j-\frac{1}{2}} & &  \\
  			a_{j-\frac{1}{2}} & 0 & a_{j-\frac{3}{2}} & \\
  			& a_{j-\frac{3}{2}} & 0 & \ddots \\
  			& & \ddots &
\ea \right) 
\,, \qquad
T_n^y =  \left( \ba{cccc} 0 & a_{j-\frac{1}{2}} & &  \\
  			-a_{j-\frac{1}{2}} & 0 & a_{j-\frac{3}{2}} & \\
  			& -a_{j-\frac{3}{2}} & 0 & \ddots \\
  			& & \ddots &
\ea \right) .
\ee 
The commutator $[T^x,T^y]= - T^z$ implies that
$2( -a^2_{j-m+\frac{1}{2}} + a^2_{j-m-\frac{1}{2}})= j-m$,
with $a_{j+\frac{1}{2}}=a_{-j-\frac{1}{2}}=0$,
from which one obtains
$a_{k}^2 = \frac{1}{4} (j-k+\frac{1}{2})(j+k+\frac{1}{2})$.
The choice of the signs of $a_k$, however, remains arbitrary.
The standard quantum mechanics convention 
(for the hermitian generators $J^i = - i T^i$) 
is to take $a_k > 0$, but we will occasionally need to
make other choices (see section \ref{SUSA}).

\setcounter{equation}{0}
\section{Some technical details} \label{appB}

Here we explain how some of the assertions made in the paper 
about vacuum solutions of the gauge theories are proved.
For the $\cN=4$ $\Sp(2N)$ (\,$\SO(N)$\,) gauge theory,
it was claimed in section \ref{Ori} that,
for a solution consisting
of a direct sum of {\it distinct} even-dimensional (odd-dimensional)
irreducible su(2) representations,
the matrix $g$ ($f$) is block-diagonal, with each block proportional
to $K$ (\ref{K}). 
In the following,  we will concentrate on the $\Sp(2N)$ case; 
the analysis for $\SO(N)$ is similar.
First, we apply the condition (\ref{spnewdef}) to $T^i$, 
where each $T^i_{n_k}$ block takes the standard form (\ref{Tz}), (\ref{Txy}). 
The equation $(T^z)^T g = -g T^z$ implies that, within each diagonal block
of $g$, 
only entries along the reverse diagonal can be non-zero.
The $T^z$ condition would permit certain non-zero entries 
in the blocks of $g$ not along the diagonal,
but these entries vanish when we impose $(T^x)^T g = -g T^x$
(assuming that the off-diagonal blocks connect representations 
of different dimensions);
thus, $g$ is block diagonal.
Finally, $(T^x)^T g = -g T^x$ implies that 
the diagonal blocks of $g$ are proportional to 
 \be
K \; = \left( \ba{cccc}  &&& 1 \\
  &&-1& \\
  &1&& \\
  \oddots &&& \\
\ea \right )\,,
\ee
provided that the signs of the $a_k$ in (\ref{Txy}) 
are chosen to satisfy $a_k = a_{-k}$ 
(the choice adopted in sections \ref{Ori} and \ref{N=2Sp}). 
The condition arising from $T^y$ 
imposes no further restrictions.
The condition $g g^*=-\1$ then implies that each block equals $K$ 
up to a phase, which may be removed by a unitary transformation.

If the direct sum contains a pair of identical su(2) representations,
then the relevant part of $g$ has the form
\be 
\left( \ba{cc}  g_{11} & g_{12} \\
                -g_{12}^T & g_{22} \ea \right ),
\ee
where the equation $(T^z)^T g = -g T^z$ implies that
only entries along the reverse diagonal of each matrix $g_{ij}$
can be non-zero.
The condition $(T^x)^T g = -g T^x$ then implies that each
$g_{ij}$ is proportional to $K$.

When we have $m$ even-dimensional representations of 
the same dimension, 
we find, repeating the above analysis pairwise 
for the constituent blocks
and using the fact that $K^T=-K$ when even-dimensional, 
that the relevant part of $g$ is equal 
(after removing some phases) 
to a real symmetric matrix tensored with $K$. 
The real symmetric matrix (tensored by $K$) 
may be diagonalized by an orthogonal matrix  (tensored by $\1$) 
that does not affect the $T^i$'s
into the form  $\diag(K,\ldots,K)$ 
(using the condition $g g^* =-\1$).
When we have $m$ odd-dimensional representations 
the analysis is similar except that, 
since $K$ is symmetric when odd-dimensional, 
one finds that the relevant part of $g$ 
is given by an antisymmetric matrix tensored with $K$. 
The condition $gg^*=-\1$ then 
implies that $m$ must be even 
(since an odd-dimensional real antisymmetric matrix 
cannot square to $-\1$), 
and one may transform $g$ into block-diagonal form, 
with each block proportional to (\ref{gform}) with $K^T=K$.

For the $\cN=2$ $\Sp(2N) + \Yasymm + 4 \Yfund$
gauge theory,
the conditions (\ref{spnewdef}) are replaced by (\ref{spantinewdef}).
For a solution consisting of a direct sum of a pair of identical 
su(2) representations,
where $g$ has the form
\be 
\left( \ba{cc}  g_{11} & g_{12} \\
                -g_{12}^T & g_{22} \ea \right ),
\ee
the equation $(T^z)^T g = -g T^z$ implies, as before, 
that only entries along the reverse diagonal of each matrix $g_{ij}$
can be non-zero.
The modified condition $(T^x)^T g =  g T^x$ then implies 
that each $g_{ij}$ is proportional, not to $K$, but to $K^\prime$ (\ref{K'}).
Since $g_{ii}$ must be antisymmetric, while $K^\prime$ is symmetric,
the diagonal blocks of $g$ must vanish, and we are left with
$g$ of the form (\ref{gnewform}). 
The generalization to the case when one has $2m$ representations 
of equal dimensions in straightforward.

\newpage

\begingroup\raggedright\endgroup


\begin{thebibliography}{10}

\bibitem{Maldacena:1998}
J.~Maldacena, ``The large $N$ limit of superconformal field theories and
  supergravity.'' Adv. Theor. Math. Phys. {\bf 2} (1998) 231--252,
  {{\tt hep-th/9711200}};
  %%CITATION = 00203,2,231;%%.
S.~S. Gubser, I.~R. Klebanov, and A.~M. Polyakov, ``Gauge theory correlators
  from non-critical string theory.'' Phys. Lett. {\bf B428} (1998) 105--114,
  {{\tt hep-th/9802109}};
  %%CITATION = PHLTA,B428,105;%%.
E.~Witten, ``Anti-de Sitter space and holography.'' Adv. Theor. Math. Phys.
  {\bf 2} (1998) 253--291, {{\tt
  hep-th/9802150}}. %%CITATION = 00203,2,253;%%.

\bibitem{Kachru:1998}
S.~Kachru and E.~Silverstein, ``4$d$ conformal theories and strings on
  orbifolds.'' Phys. Rev. Lett. {\bf 80} (1998) 4855--4858,
  {{\tt hep-th/9802183}};
  %%CITATION = PRLTA,80,4855;%%.
A.~Lawrence, N.~Nekrasov, and C.~Vafa, ``On conformal field theories in four
  dimensions.'' Nucl. Phys. {\bf B533} (1998) 199--209,
  {{\tt hep-th/9803015}};
  %%CITATION = NUPHA,B533,199;%%.
Y.~Oz and J.~Terning, ``Orbifolds of $AdS_5\times S_5$ and 4$d$ conformal field
  theories.'' Nucl. Phys. {\bf B532} (1998) 163,
  {{\tt hep-th/9803167}};
  %%CITATION = NUPHA,B532,163;%%.
S.~Gukov, ``Comments on $\cN = 2$ AdS orbifolds.'' Phys. Lett. {\bf B439}
  (1998) 23--28, {{\tt
  hep-th/9806180}}. %%CITATION = PHLTA,B439,23;%%.

\bibitem{Kakushadze:1998a}
Z.~Kakushadze, ``Gauge theories from orientifolds and large $N$ limit.'' Nucl.
  Phys. {\bf B529} (1998) 157--179,
  {{\tt hep-th/9803214}};
  %%CITATION = NUPHA,B529,157;%%.
Z.~Kakushadze, ``On large $N$ gauge theories from orientifolds.'' Phys. Rev.
  {\bf D58} (1998) 106003, {{\tt
  hep-th/9804184}}. %%CITATION = PHRVA,D58,106003;%%.

\bibitem{Fayyazuddin:1998}
A.~Fayyazuddin and M.~Spalinski, ``Large $N$ superconformal gauge theories and
  supergravity orientifolds.'' Nucl. Phys. {\bf B535} (1998) 219--232,
  {{\tt hep-th/9805096}}.
  %%CITATION = NUPHA,B535,219;%%.

\bibitem{Witten:1998c}
E.~Witten, ``Baryons and branes in anti de Sitter space.'' JHEP {\bf 07} (1998)
  006, {{\tt hep-th/9805112}}.
  %%CITATION = JHEPA,9807,006;%%.

\bibitem{Aharony:1998}
O.~Aharony, A.~Fayyazuddin, and J.~Maldacena, ``The large $N$ limit of $\cN =
  2,1$ field theories from three-branes in F-theory.'' JHEP {\bf 07} (1998)
  013, {{\tt hep-th/9806159}}.
  %%CITATION = JHEPA,9807,013;%%.

\bibitem{Gukov:1998b}
S.~Gukov and A.~Kapustin, ``New $\cN = 2$ superconformal field theories from
  M/F theory orbifolds.'' Nucl. Phys. {\bf B545} (1999) 283--308,
  {{\tt hep-th/9808175}}.
  %%CITATION = NUPHA,B545,283;%%.

\bibitem{Lozano:2000}
I.~P. Ennes, C.~Lozano, S.~G. Naculich, and H.~J. Schnitzer, ``Elliptic models,
  type IIB orientifolds and the AdS/CFT correspondence.'' Nucl. Phys. {\bf
  B591} (2000) 195--226, {{\tt
  hep-th/0006140}}. %%CITATION = NUPHA,B591,195;%%.

\bibitem{Girardello:1999}
L.~Girardello, M.~Petrini, M.~Porrati, and A.~Zaffaroni, ``The supergravity
  dual of $\cN = 1$ super Yang-Mills theory.'' Nucl. Phys. {\bf B569} (2000)
  451--469, {{\tt
  hep-th/9909047}}. %%CITATION = NUPHA,B569,451;%%.

\bibitem{Polchinski:2000}
J.~Polchinski and M.~J. Strassler, ``The string dual of a confining
  four-dimensional gauge theory.''
  {{\tt hep-th/0003136}}.
  %%CITATION = HEP-TH 0003136;%%.

\bibitem{Klebanov:1998}
I.~R. Klebanov and E.~Witten, ``Superconformal field theory on threebranes at a
  Calabi-Yau singularity.'' Nucl. Phys. {\bf B536} (1998) 199,
  {{\tt hep-th/9807080}};  
  %%CITATION = NUPHA,B536,199;%%.
I.~R. Klebanov and M.~J. Strassler, ``Supergravity and a confining gauge
  theory: Duality cascades and ($\chi$)SB-resolution of naked singularities.''
  JHEP {\bf 08} (2000) 052, {{\tt
  hep-th/0007191}};  
%%CITATION = JHEPA,0008,052;%%. 
K.~Pilch and N.~P.~Warner,
``$\cN = 1$ supersymmetric renormalization group flows from IIB supergravity,''
{{\tt hep-th/0006066}}; 
%%CITATION = HEP-TH 0006066;%%
J.~M. Maldacena and C.~Nunez, ``Towards the large $N$ limit of pure $\cN = 1$
  super Yang Mills.'' {{\tt
  hep-th/0008001}}; 
%%CITATION = HEP-TH 0008001;%%.
C.~Vafa, ``Superstrings and topological strings at large $N$.''
{{\tt hep-th/0008142}}; 
%%CITATION = HEP-TH 0008142;%%
M.~Gra\~{n}a and J.~Polchinski,
``Supersymmetric three-form flux perturbations on $\mathrm{AdS}_5$.''
Phys. Rev. D {\bf 63} (2001) 026001, {\tt hep-th/0009211}.
%%CITATION = HEP-TH 0009211;%%

\bibitem{Aharony:2000}
O.~Aharony and A.~Rajaraman, ``String theory duals for mass-deformed SO($N$)
  and USp($2N$) $\cN = 4$ SYM theories.'' Phys. Rev. {\bf D62} (2000) 106002,
  {{\tt hep-th/0004151}}.
  %%CITATION = PHRVA,D62,106002;%%.

\bibitem{Vafa:1994}
C.~Vafa and E.~Witten, ``A strong coupling test of S duality.'' Nucl. Phys.
  {\bf B431} (1994) 3--77, {{\tt
  hep-th/9408074}};  
%%CITATION = NUPHA,B431,3;%%.
R.~Donagi and E.~Witten, ``Supersymmetric Yang-Mills theory and integrable
  systems.'' Nucl. Phys. {\bf B460} (1996) 299--334,
  {{\tt hep-th/9510101}}.
  %%CITATION = NUPHA,B460,299;%%.

\bibitem{Myers:1999}
R.~C. Myers, ``Dielectric-branes.'' JHEP {\bf 12} (1999) 022,
  {{\tt hep-th/9910053}}.
  %%CITATION = JHEPA,9912,022;%%.

\bibitem{Kac:1999}
V.~G. Kac and A.~V. Smilga, ``Normalized vacuum states in $\cN = 4$
  supersymmetric Yang-Mills quantum mechanics with any gauge group.'' Nucl.
  Phys. {\bf B571} (2000) 515--554,
  {{\tt hep-th/9908096}}.
  %%CITATION = NUPHA,B571,515;%%.

\bibitem{Kabat:1998}
D.~Kabat and W.~{Taylor IV}, ``Linearized supergravity from matrix theory.''
  Phys. Lett. {\bf B426} (1998) 297--305,
  {{\tt hep-th/9712185}}.
  %%CITATION = PHLTA,B426,297;%%.

\bibitem{Hanany:1999}
A.~Hanany, B.~Kol, and A.~Rajaraman, ``Orientifold points in M theory.'' JHEP
  {\bf 10} (1999) 027, {{\tt
  hep-th/9909028}}. %%CITATION = JHEPA,9910,027;%%.

\bibitem{Gimon:1996a}
E.~G. Gimon and J.~Polchinski, ``Consistency conditions for orientifolds and
  D-manifolds.'' Phys. Rev. {\bf D54} (1996) 1667--1676,
  {{\tt hep-th/9601038}}.
  %%CITATION = PHRVA,D54,1667;%%.

\bibitem{Sen:1996}
A.~Sen, ``F-theory and orientifolds.'' Nucl. Phys. {\bf B475} (1996) 562--578,
  {{\tt hep-th/9605150}}; 
  %%CITATION = NUPHA,B475,562;%%.
T.~Banks, M.~R. Douglas, and N.~Seiberg, ``Probing F-theory with branes.''
  Phys. Lett. {\bf B387} (1996) 278--281,
  {{\tt hep-th/9605199}}; 
  %%CITATION = PHLTA,B387,278;%%.
O.~Aharony, J.~Sonnenschein, S.~Yankielowicz, and S.~Theisen, ``Field theory
  questions for string theory answers.'' Nucl. Phys. {\bf B493} (1997)
  177--197, {{\tt
  hep-th/9611222}};  
%%CITATION = NUPHA,B493,177;%%.
M.~R. Douglas, D.~A. Lowe, and J.~H. Schwarz, ``Probing F-theory with multiple
  branes.'' Phys. Lett. {\bf B394} (1997) 297--301,
  {{\tt hep-th/9612062}}.
  %%CITATION = PHLTA,B394,297;%%.

\bibitem{Park:1998}
J.~Park and A.~M. Uranga, ``A note on superconformal $\cN = 2$ theories and
  orientifolds.'' Nucl. Phys. {\bf B542} (1999) 139--156,
  {{\tt hep-th/9808161}}.
  %%CITATION = NUPHA,B542,139;%%.

\bibitem{Erlich:1999}
J.~Erlich, A.~Hanany, and A.~Naqvi, ``Marginal deformations from branes.'' JHEP
  {\bf 03} (1999) 008, {{\tt
  hep-th/9902118}}. %%CITATION = JHEPA,9903,008;%%.

\bibitem{Witten:1998a}
E.~Witten, ``Toroidal compactification without vector structure.'' JHEP {\bf
  02} (1998) 006, {{\tt
  hep-th/9712028}}. %%CITATION = JHEPA,9802,006;%%.

\bibitem{Douglas:1996}
M.~R. Douglas and G.~Moore, ``D-branes, Quivers, and ALE Instantons.''
  {{\tt hep-th/9603167}}; 
  %%CITATION = HEP-TH 9603167;%%.
M.~R. Douglas, B.~R. Greene, and D.~R. Morrison, ``Orbifold resolution by
  D-branes.'' Nucl. Phys. {\bf B506} (1997) 84--106,
  {{\tt hep-th/9704151}}.
  %%CITATION = NUPHA,B506,84;%%.

\end{thebibliography}
\end{document}